\documentclass{nature}

\usepackage{amssymb}
\usepackage{amsmath}
\usepackage{wasysym}
\usepackage{graphicx}
\usepackage{xspace}
\usepackage[usenames, dvipsnames]{color}
\usepackage[normalem]{ulem}
\usepackage{soul}
\usepackage[displaymath,mathlines]{lineno}
\usepackage{subfig}
\usepackage[font=footnotesize]{caption}

\usepackage{floatrow}
\DeclareFloatFont{tiny}{\scriptsize}

\newcommand\apj{Astrophys. J.}
\newcommand\apjl{Astrophys. J. Lett.}
\newcommand\apjs{Astrophys. J., Suppl.}
\newcommand\mnras{mnras}
\newcommand\nat{Nature}
\newcommand\physrep{Phys. Rep.}

\usepackage{xpatch}

\newcommand{\aap}{Astron. Astrophys.}

\newcommand{\ssr}{S. Science Rev.}
\newcommand{\jcap}{J. Cosm. and Astro. Phys.}
\newcommand{\iaucirc}{IAU Circ.}
\newcommand{\aj}{Astrophys. J.}
\newcommand{\na}{N. Astro.}

\title{Relativistic Effects and GRB Polarization in Power-Law Evolution}

\newcounter{firstbib}

\begin{document}

\maketitle

\author{
Liang~Li$^{\ref{ICRANet},\ref{INAF},\ref{ICRA}}$,
She-Sheng~Xue$^{\ref{ICRANet},\ref{INAF},\ref{ICRA}}$,
and Zi-Gao ~Dai$^{\ref{USTC},\ref{NU}}$
}

\begin{affiliations}
\small 
\item ICRANet, Piazza della Repubblica 10, 65122 Pescara, Italy \label{ICRANet}
\item INAF -- Osservatorio Astronomico d'Abruzzo, Via M. Maggini snc, I-64100, Teramo, Italy \label{INAF}
\item Dip. di Fisica and ICRA, Sapienza Universita di Roma, Piazzale Aldo Moro 5, I-00185 Rome, Italy \label{ICRA}
\item Department of Astronomy, University of Science and Technology of China, Hefei 230026, China \label{USTC}
\item School of Astronomy and Space Science, Nanjing University, Nanjing 210023, China \label{NU}
\end{affiliations}

\begin{abstract}

Despite decades of polarization observations and high-significance polarized $\gamma$-ray, X-ray, optical, and radio emissions in gamma-ray bursts (GRBs) have been accumulating in dozens of cases\cite{Frail1998GCN,Covino1999,Coburn2003,Mundell2007Sci,Steele2009,Zhang2019,Chattopadhyay2019}, people has yet to find a consistent scenario for understanding the globally observed timing properties of GRB polarization to date. Here, 
we report that the observed properties of GRB polarization exhibit a four-segment timing evolution at the cosmological distance: (I) an initial hump early on (within the first few seconds); (II) a later on power-law decay (from $\sim$10$^{1}$ to $\sim$10$^{4}$ s), which takes the form of $\pi_{\rm obs} \propto t^{-0.50 \pm 0.02}$; (III) afterwards a late-time rebrightening hump (from $\sim$10$^{4}$ to $\sim$10$^{5}$ s); and (IV) finally a flatting power-law decay (from $\sim$ 10$^{5}$ to $\sim$ 10$^{7}$ s), with the the form of $\pi_{\rm obs} \propto t^{-0.21 \pm 0.08}$. These finding may present a challenge to the mainstream of polarization models that assume the polarization time evolution change in different emission regions. We show that these results can be explained by relativistic and geometric effects\cite{Fan2008} of a highly relativistic and magnetized jet generated by central engine, and  ``magnetic patches''\cite{Gruzinov1999,Granot2003,Nakar2004,Granot2005} distributed as a globally random but locally coherent form. The long-term timing evolution of observed GRB polarization follows a scaling law $\pi_{\rm obs}\propto 1/S_{\rm obs}$, dominantly determined by how ``magnetic patches'' are randomly distributed in observed emission region $S_{\rm obs}$ on the jet plane of $1/\Gamma$ cone. 
It predicts the polarization hump and tail form in accordance with the luminosity jet break phenomenon. Our analysis suggests that there is a single dominant mechanism (relativistic and geometric effects) that might account for the global observational properties of GRB polarization, and other emission mechanisms and effects might play a role in spatially local and temporally short effects on GRB polarization.

\end{abstract}

Measuring the polarization from GRB emission play a crucial role in understanding the physics of relativistic jets--in particular--providing a unique probe of the magnetic fields in the collimated jets\cite{Usov1992,Medvedev1999,Piran2005,Metzger2011,Zhang2011b,Pudritz2012}. After more than two decades of polarimetric observations, GRB polarimetric measurements have accumulated in dozens of cases\cite{Covino1999,Mundell2007Sci,Steele2009}, and have been observed in both early and late emission phases, spanning from a few seconds to a few months (covering six to seven orders of magnitude in time) after the burst trigger, making it possible to study GRB polarization properties in a global manner. 

Despite decades of multi-wavelength observations and considerable theoretical efforts, there is still no a consistent scenario for explaining the globally observed timing properties of GRB polarization to date. A better understanding of the physics of GRB polarization properties is required to have a more complete picture of the long-timescale evolutionary connections from burst to burst. Therefore, we searched the literature for published and archival observations that would allow us to provide a complete picture of polarization evolution. With this dedicated search, the complete GRB polarization sample, which
consists of 73 bursts (39 bursts with known redshifts) and covers a broad wavelength range (from radio to $\gamma$-ray) of polarization measurements, is provided (see Table \ref{tab:fullsample}).

In order to investigate the temporal properties of GRB polarization, a corresponding epoch of the polarization measurement for each burst is required. In order to explore the intrinsic properties of GRB polarization at the cosmological distance, a time-dilation correction factor 1/(1+$z$) is applied. The degree of polarization ($\pi_{\rm obs}$) as a function of time after the burst plotted in the cosmological rest-frame is shown in Figure \ref{fig:piTimefull} for the full polarization sample (with known redshift, see Table \ref{tab:fullsample}).
By modeling the data, we discover that the globally time-sampled evolutionary properties of GRB polarization degrees $\pi_{\rm obs}$ (percentages, ranging from 0\% to 100\%) consist of four segments (Figure \ref{fig:piTimefull}): (I) an initial hump early on (within the first few seconds); (II) a power-law decay over a long-term timescale period (from $\sim$10$^{1}$ to $\sim$10$^{4}$ s) later on, which takes the form of $\pi_{\rm obs} \propto t^{-0.50 \pm 0.02}$ with a decay index around at -0.50 (see Methods); (III) a rebrightening hump occurs at a later time (from $\sim$10$^{4}$ to $\sim$10$^{5}$ s); and (IV) finally, a new power-law decay (from $\sim$ 10$^{5}$ to $\sim$ 10$^{7}$ s), which takes the form of $\pi_{\rm obs} \propto t^{-0.21 \pm 0.08}$. 
These observational properties are naturally in agreement with theoretical previsions based on the scenario of relativistic geometric effects and outflow dynamics (Figure \ref{fig:jet}).
In the standard GRB fireball ``internal-external'' shock model\cite{Rees1994,Meszaros1997,Sari1998,Dai1998,Piran1999,Chevalier2000}, following the collimated relativistic outflows (jets) launched from the central engine: the early and short-lived prompt $\gamma$-ray emissions generated by an internal shock, where two consecutive shells interact and collide, occur at a close distance from the progenitor; while the late and long-lasting, X-ray, optical, and radio wavelengths afterglow emissions originated from an external shock, where outflows interact with ambient medium, occur at large distances from progenitors. Within this scenario, relativistic magnetized jets are also expected to be generated by the central engines and launched along with the relativistic collimated jets, either by hyper-accreting black holes or by rapidly spinning magnetars\cite{Blandford1977,Usov1992,Thompson1994,Medvedev1999,Spruit2001,Lyutikov2003,Piran2005,Dai2006,Metzger2011,Ruffini2010,Zhang2011b,Pudritz2012,Li2018b,Xue2021}, very likely distributed as globally random, but locally coherent magnetic fields. A local coherent magnetic domain of typical area dimension $S_{\rm patch}$ is named as a ``magnetic patch''\cite{Gruzinov1999,Granot2003,Nakar2004,Granot2005}, where  magnetic field $H_{\rm patch}$ orientation is in order. It emits polarized photon about $\sim \Pi_{\rm max}$ (Eq.\ref{eq:Pi_max}). The orientations of magnetic fields and photon polarization vary from one magnetic patch to another. These magnetic patches and fields randomly distribute in a outflow due to magnetized fluid dynamics. 

In early emission (within the first few seconds), starting at an bulk Lorentz factor $\Gamma\gg 1$, due to relativistic beaming, only the emission inside the 1/$\Gamma$ cone can be contributed to the observed flux\cite{Rybicki1979}, so as to the observable area $S_{\rm obs}$ is a small portion of outflow. When only one magnetic patch in outflow is seen ($S_{\rm obs}\lesssim S_{\rm patch}$), the observed polarization $\pi_{\rm obs}$ is about $\Pi_{\rm max}$, associating to magnetic patch field $H_{\rm patch}$ and orientation. Afterwards, $\Gamma$ decreases due to energy dissipation, and $S_{\rm obs}$ increases. More and more magnetic patches in outflow are seen. The observed net polarization from all magnetic patches seen should be smaller than $\Pi_{\rm max}$, attributed to globally and randomly disorder distributions of magnetic patch fields and orientation in a outflow. The more magnetic patches are seen and contribute to the observed flux, the smaller measured net polarization is, owing to an increase of randomness degree of magnetic patches distributing over the observed area $S_{\rm obs}$. Considering that the local synchrotron radiation and magnetized fluid dynamics timescales are much smaller than the period of observed polarization data collected and analyzed, we propose that outflow relativistic and geometric effects play an essential role in explaining observed polarization properties in a long-time duration. The observed polarization $\pi_{\rm obs}$ should correspond to the net polarization averaged over magnetic patches distribution and their fluid and radiation dynamics effects. As a consequence, the $\pi_{\rm obs}$ long-time evolution mainly follows a scaling law $\pi_{\rm obs}\propto 1/S_{\rm obs}$ ($S_{\rm patch} \lesssim S_{\rm obs} \lesssim S_{\rm jet}$ and from $\sim 10^{1}$ to $\sim 10^{4}$ s) until a jet-break time (from $\sim 10^4$ to $\sim 10^5$ s) is reached when $S_{\rm obs}\approx S_{\rm jet}$ (see Methods). The scaling law seems in agreement with data. At the moment of jet-break time, an Earth based observer naturally starts to see non-axial symmetric or irregular boundaries of outflows. Such an irregular or fractal geometry of outflow boundary decrease the randomness of spatial distributions of magnetic patches, and should in principle result in net polarization increases. Therefore the scenario {\it a priori} previews that at jet-break time the observed polarization has an increment correlating known afterglow jet break phenomena, and change its decay scaling law from $\pi_{\rm obs}\propto 1/S_{\rm obs}$ to a shallower one, since total net polarization of entire jet is seen and the observable area $S_{\rm obs} >S_{\rm jet}$ becomes irrelevant in net polarization time evolution. In four GRBs polarization data (see Table \ref{tab:JetBreak}), we indeed find around $10^{4}$ to $10^{5}$ seconds bit-increments appear correlating light-curve jet breaks, and a shallower decaying law follows after $10^{5}$ seconds, the typical epochs and temporal slopes for jet breaks that are observed for these bursts (see Methods). We also constrain shallower decaying law in polarization data after the jet-break time ($S_{\rm obs}> S_{\rm jet}$ and $\gtrsim 10^{5}$ seconds), see $\Gamma \gtrsim 1$ in Figure \ref{fig:jet}, the observed net polarization decaying law clearly deviates from the scaling power law $\pi_{\rm obs}\propto 1/S_{\rm obs}$ and becomes shallower one. It may remain as a constant soon after the jet-break time, then decreases slowly, due to the time variations of magnetic patch fields and sizes, as well as magnetized outflow itself after jet break.

There are two remarks in order. First, in the first few seconds (Segment I, see Figure \ref{fig:piTimefull}) for the early polarization evolution, the observed area should be smaller than a magnetic patch area $S_{\rm obs}< S_{\rm patch}$ because of a giant $\Gamma$, and the observed polarization $\pi_{\rm obs}\sim \Pi_{\rm max}(H_{\rm patch})$ should follow the variations of magnetic patch field $H_{\rm patch}$ and electron distribution. We speculate $\pi_{\rm obs}\sim \Pi_{\rm max}$ slightly increases, reaching its maximal value when $S_{\rm obs}\approx S_{\rm patch}$, for the possible reasons that $\Gamma$ decreases, $S_{\rm obs}$ increases and more electrons (involved in) emitting radiation are observed in time. Unfortunately, current observations do not contain high-resolution polarization data for this earliest period, and we have not been able to determine the polarization evolution in time. Second, in the period after the jet break $>10^5$ sec and $S_{\rm obs}> S_{\rm jet}$ (Segment IV, see Figure \ref{fig:jet}) for the later polarization evolution, we are lack of better-established polarization data. Therefore  the power-law decay index during Segment IV is poorly constrained as determined by the current sample. To gain a deeper insight into the polarization time evolution in early (Segment I) and later (Segment IV) periods, we look forward to high-sensitivity polarimetric observations from upcoming instruments\cite{DeAngelis2021} to establish well-sampled polarization light curves in these two periods. It will be subjects for future studies. 

In addition to the time-sampled polarization lightcurve from observed data, we also discuss other observational evidences supporting the scaling law in Segment II and bit-increment in Segment III. In Segment II ($S_{\rm patch} \lesssim S_{\rm obs} < S_{\rm jet}$), the observed net polarization mainly follows a scaling law $\pi_{\rm obs}\propto S^{-1}_{\rm obs}$ and $S_{\rm jet}/S_{\rm obs}=\theta^{2}_j/\theta^{2}_e$ (see Methods). Due to relativistic beaming, only the emission inside the 1/$\Gamma$ cone contributes to the observed flux, one has $\theta_{e}\simeq \Gamma^{-1}$. One therefore has $\pi_{\rm obs} \propto \Gamma^{2}$. A possible test can be made by investigating the relation between the polarization data $\pi_{\rm obs}$ and Lorentz factor $\Gamma(\pi)$ and the latter is measured in the same epoch of polarization data (see Methods). Since $\Gamma(\pi)$ cannot be precisely estimated for certain bursts, one can use $\Gamma_{0}$ as a proxy. An empirical relation\cite{Lv2012,Liang2010} ($\Gamma_{0}\simeq 249 L^{0.30}_{\gamma,\rm iso,52}$) is used to estimate the values of $\Gamma_{0}$, as long as their energy flux $F_{\gamma}$ and redshift are known (see Methods). Extended Data Figure \ref{fig:piGamma} shows $\pi_{\rm obs}$ as functions of $\Gamma_{0}$. We find the data points indeed cluster around the scaling law $\pi_{\rm obs} \propto \Gamma^{2}$ (see the solid lines of Figure \ref{fig:piGamma}), where various external shock models are considered. Moreover, in the epoch of jet-break (Segment III, $S_{\rm obs}\approx S_{\rm jet}$), we find five bursts (GRB 990510, GRB 010222, GRB 020405, GRB 030328, and GRB 080928) that exhibit the bit-increment polarization data. It is further shown that in these bursts the occurrences of polarization bit-increments are in agreement with the jet-break analysis using their afterglow data, where three (or possibly four) bursts may observe a jet break (see Methods).

The intrinsic mechanisms and effects that account for GRB polarization are complicated. The observed net polarization should come from the different microscopic radiation mechanisms and macroscopic effects of jet geometries, viewing angles, magnetic field configurations, and degree of magnetized jet, etc. Our data and theoretical analysis suggest that the relativistic geometrical effect is the dominant one across various emission regions, naturally explaining the observed power-law evolution of GRB polarization in a long time scale ${\mathcal O}(10^6)$ seconds. This single power-law over long time scales across multiple emission regions challenges the current mainstream of polarization models that assume different polarization properties in distinct emission regions\cite{Granot2003Natur,Sagiv2004,Lundman2018,Zhang2011b}, as described by these models, the temporal evolution of GRB polarization may display a ``three-stage jump" pattern (see Figure \ref{fig:com_model}) from early prompt emission to later afterglow ``reverse-forward" shock emission regions. These results suggest that other possible mechanisms and effects (e.g., the internal/external shock model\cite{Rees1994}, the dissipative photosphere model\cite{Rees2005,Peer2006a}, the ICMART model\cite{Zhang2011b}) give contributions to GRB polarization in small time scales, that result in observational data scattering up and down around the power-law evolution.

We find four dramatically different segments spanning from a few seconds to a few months, covering six to seven orders of magnitude in time. They characterize a complete picture of the global timing properties of GRB polarization for the first time. Our results show that the time evolution of GRB polarization follows the  power-law decay with an index $\alpha \simeq -0.50$ until the jet-break time. It may provide an empirical estimate of $\pi_{\rm obs}$ at a given time. The estimated $\pi_{\rm obs}$ value can be used for further constraints on physical models and their relevant parameters. For instance, the scaling law $\pi_{\rm obs}\propto S^{-1}_{\rm obs}\propto \Gamma^{2}$ gives an insight into the traditional GRB afterglow models, which describe the environment surrounding GRBs (see Figure \ref{fig:piTimeModel} and Methods). Our result supports the afterglow scenarios of constant energy and density stratification (wind) environment\cite{Dai1998,Chevalier2000}. It may disfavor the case of a large-scale ordered magnetic field advocated from the central engine with an on-axis observation, because an rising scaling law of GRB polarization is predicted by the case; but favor the case of a large-scale ordered magnetic field with an off-axis observation. Moreover, analogously to the Amati relation\cite{Amati2002}, the GRB polairzation power law evolution in time shows though observed GRB phenomena are individually different (e.g., total energy, lightcurve, and spectrum), they have some common features in central engine and relativistic magnetized jet evolution at large scales in time and space.
 

 --------------------------------------------------------------
 



\section*{Author contributions}
LL led the data analysis, and contributed to part of the physical explanations. SSX proposed the theoretical model for explanations and previsions of the observational data in four-time segments. LL and SSX wrote the article. ZGD participated in discussions. All authors have reviewed, discussed, and commented on the present results and on the manuscript.

\section*{Acknowledgements}
LL thank Yu~Wang, Felix~Ryde, Bing~Zhang, Xue-Feng~Wu, Jin-Jun~Geng, Soroush Shakeri, Shuang-Nan~Zhang, Remo~Ruffini, and the ICRANet members for many helpful discussions on GRB physics and phenomena. 

\section*{Author information}
Correspondence and requests for materials should be addressed to LL (liang.li@icranet.org), SSX (xue@icra.it), and ZGD (dzg@nju.edu.cn).

\section*{Competing Interests}
The authors declare that they have no competing financial interests.


\clearpage
\begin{figure}[ht!]
\includegraphics[width=1\textwidth]{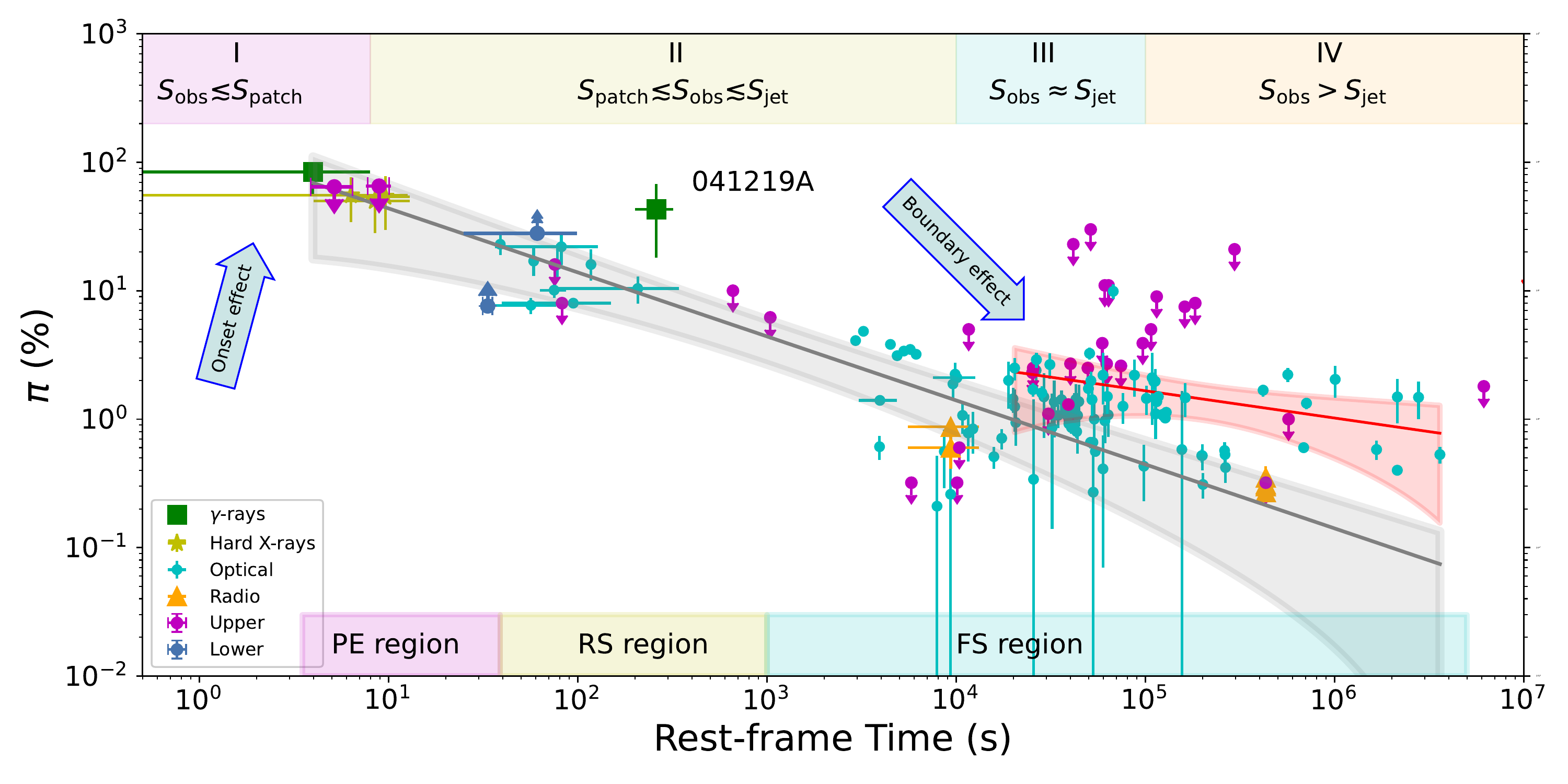}
\caption{The degree of polarization ($\pi_{\rm obs}$, percentages ranging from 0\% to 100\%) for the full sample with known redshift (see Table \ref{tab:fullsample}) is plotted as a function of time after the burst at the cosmological distance. Data points with solid colors represent the polarization data in different energy band. Green: $\gamma$-rays; yellow: hard X-rays; cyan: optical; orange: radio; blue-magenta: the polarization data placed at the upper limits on the measurements; red-green: the polarization data placed at the lower limits on the measurements. Segment I: an initially slight hump caused by the onset effect ($S_{\rm obs}\lesssim S_{\rm patch}$) within the first few seconds. Segment II: a power-law decay ($S_{\rm patch} \lesssim S_{\rm obs} \lesssim S_{\rm jet}$) over a long timescale period (from $\sim$10$^{1}$ to $\sim$10$^{4}$ s), which takes the form of $\pi_{\rm obs} \propto t^{-0.50 \pm 0.02}$. Segment III: a late-rebrightening hump caused by the ``jet break'' boundary effect ($S_{\rm obs}\approx S_{\rm jet}$) occurs at a later time (from $\sim$10$^{4}$ to $\sim$10$^{5}$ s). Segment IV: a flatting power-law decay ($S_{\rm obs}> S_{\rm jet}$) after the ``jet break'' (from $\sim$ 10$^{5}$ to $\sim$ 10$^{7}$ s), which takes the form of $\pi_{\rm obs} \propto t^{-0.21 \pm 0.08}$. The solid line is the best fits using the power-law model and with $2\sigma$ (95\% confidence interval) error shadow region.}
\label{fig:piTimefull}
\end{figure}

\clearpage
\begin{figure*}
\includegraphics[angle=0,scale=0.45]{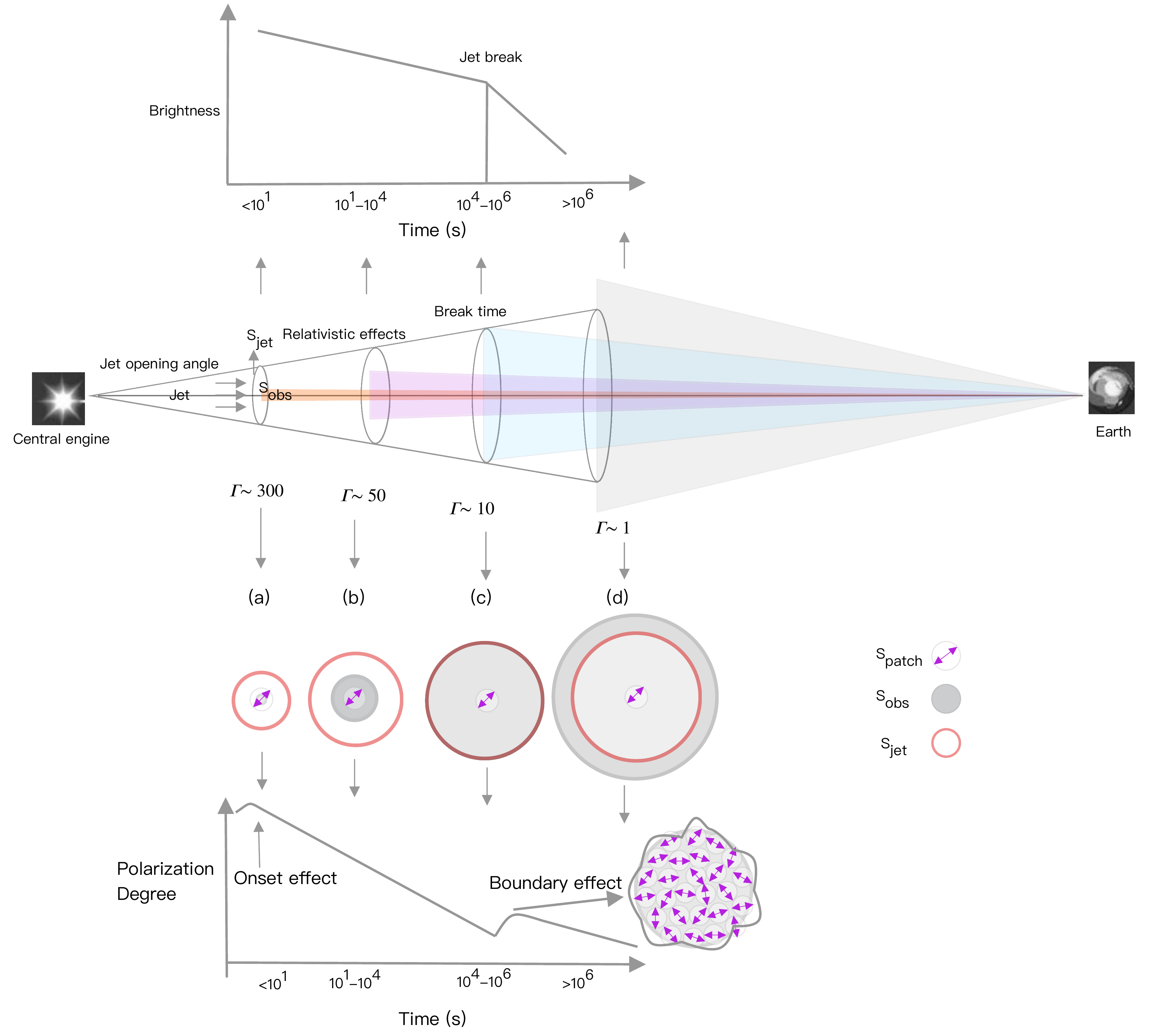}
\caption{A cartoon picture for depicting the timing evolution of the GRB polarization degree based on relativistic geometrical effects. For simplicity, we assume that the jet opening angle $\theta_{j}$ remains unchanged throughout jet evolution, corresponding to an area projected onto the jet plane as $S_{\rm jet}$ (see white circle symbol in the right panel). Due to relativistic beaming, only the emission inside the 1/$\Gamma$ cone contributes to the observed flux, corresponding to an observed area projected onto the jet plane as $S_{\rm obs}$ (see grey circle symbol in the right panel). A ``magnetic patch'' corresponds to an area projected onto the jet plane as $S_{\rm patch}$ (see orange circle symbol in the right panel). (a) The onset effect (Segment I) occurs at an early emission (within the first few seconds), corresponding to $S_{\rm obs}\lesssim S_{\rm patch}$ with a huge $\Gamma\gg 1$ (e.g., $\Gamma$=300). (b) The main power-law decay phase (Segment II) occur over a long timescale period (from $\sim$ 10$^{1}$ to $\sim$ 10$^{4}$ s), corresponding to $S_{\rm patch} \lesssim S_{\rm obs} \lesssim S_{\rm jet}$ with a moderate $\Gamma$ (e.g., $\Gamma$=50). (c) The ``jet break'' boundary effect (Segment III) occurs at a later emission when the ``jet break'' is reached (from $\sim$ 10$^{4}$ to $\sim$ 10$^{5}$ s), corresponding to $S_{\rm obs}\approx S_{\rm jet}$ with a relatively low $\Gamma$ (e.g., $\Gamma$=10). (d) Finally the new power-law decay phase (Segment IV) occurs after the ``jet break'' (from $\sim$ 10$^{5}$ to $\sim$ 10$^{7}$ s), corresponding to $S_{\rm obs}> S_{\rm jet}$ with a small $\Gamma$ (e.g., $\Gamma$=1).}\label{fig:jet}
\end{figure*}

\clearpage
\begin{figure}[ht!]
\includegraphics[width=0.6\textwidth]{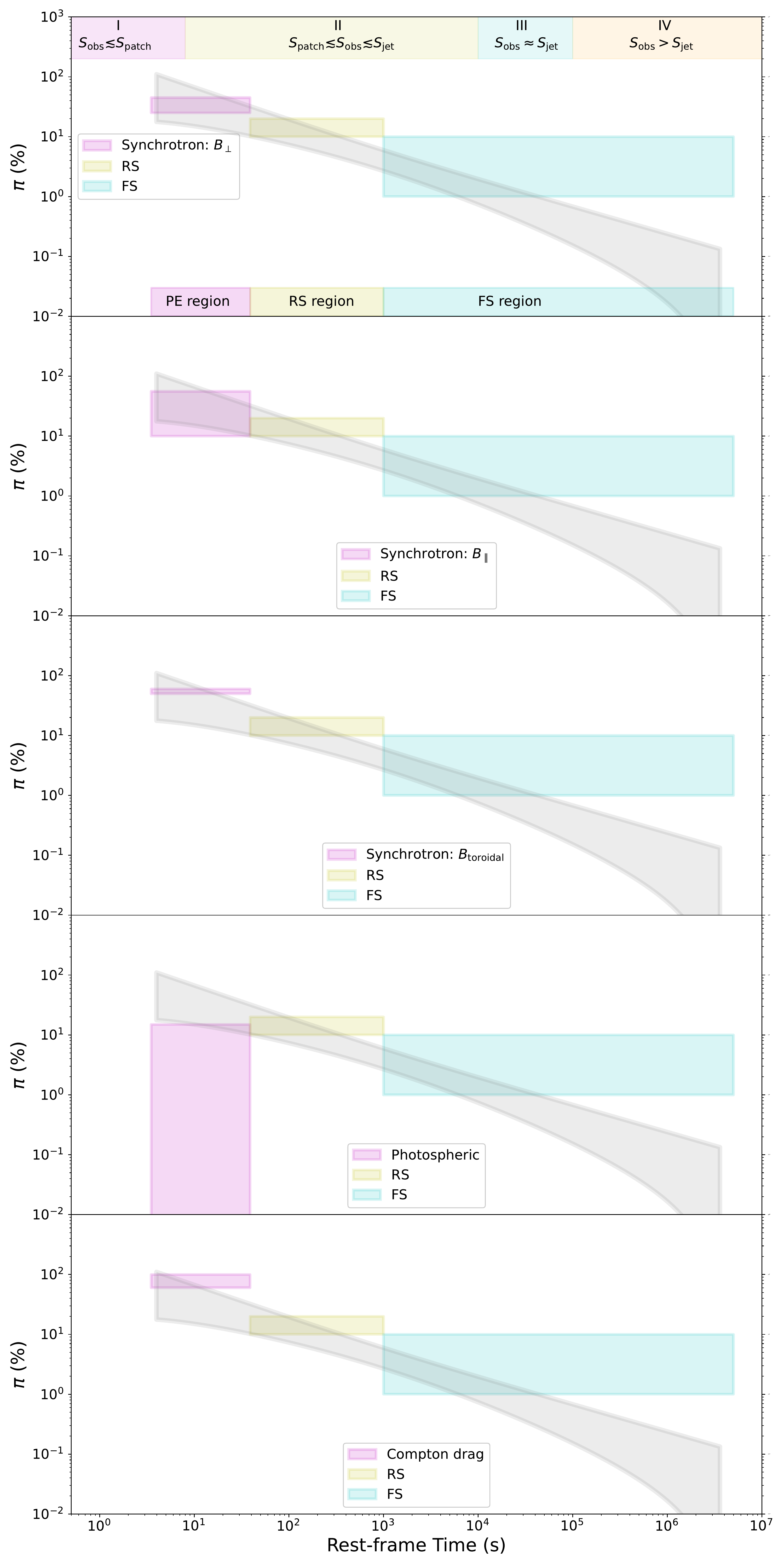}
\caption{The contrast between the observed polarization degree and the theoretical predictions by different radiation mechanisms\cite{2020MNRAS.491.3343G}. In the prompt emission (PE) phase, we consider the synchrotron radiation from different magnetic field structures: (a) a random field normal to the radial direction $B_{\perp}$ ($\pi \sim$ 25\%-45\%); (b) an ordered field radial $B_{\parallel}$ ($\pi \sim$ 10\%-56\%); (c) a toroidal $B_{\rm tor}$ ($\pi \sim$ 50\%-60\%), as well as (d) photosphere emission model ($\pi \sim$ 0\%-15\%); (e) Compton drag model\cite{Lazzati2004} ($\pi \sim$ 60\%-100\%). In the afterglow emission phase, we consider the highly polarized reverse-shock (RS,\cite{ZhangShuai2015,2013Natur.504..119M}) emission ($\pi \sim$ 10\%-20\%), and forward-shock (FS,\cite{Gruzinov1999}) emission ($\pi \sim$ 1\%-10\%).}
\label{fig:com_model}
\end{figure}

\newpage
\clearpage
\begin{center}
\textbf{Methods}
\end{center}

\subsection{The GRB polarimetric sample} \label{sec:Sample}

We conducted an extensive search of the literature for published and archival polarimetric observations, and attempted to include all the bursts that had polarization measurements to date. With this dedicated search, the complete GRB polarization sample, which consists of 73 bursts (39 bursts with known redshifts, see Figure \ref{fig:piTimefull}) and covers a broad wavelength range (from radio to $\gamma$-ray emission) of polarization measurements, is provided (see Table \ref{tab:fullsample}).

These observed polarization data are categorized into three types of data sets: i) Time-integrated polarization data, which represent average polarimetric properties, for the entire emission period are treated as a single time event, resembling the approach used in GRB spectral analyses; ii) Time-resolved polarization data, in which the whole emission period is divided into multiple timing events, and polarimetric measurements, are therefore performed on each event individually; iii) Polarimetric measurements are placed on an upper/lower limit. We note that (1) For a given burst, time-integrated polarization data can in principle include as many numbers of detected photons as possible, thereby reducing the uncertainties and avoiding unreliable fluctuations introduced by the time-resolved data; (2) The polarization data for a good fraction of the bursts have placed an upper/lower limit on the measurements, and therefore, this data set may not be as tightly constrained (has a loose constraint) to the model as on the other data sets. 

\subsection{Modeling the multi-wavelength time-sampled polarization data}

In order to explore the timing properties of GRB polarization, a corresponding epoch of the polarization measurement for each burst is required. To investigate the intrinsic properties of GRB polarization, we select those bursts whose redshift measurements are also reported. Extended Data Figure \ref{fig:piTimeInterval} shows that the time-integrated polarization degree $\pi_{\rm obs}$ (percentage, ranges from 0\% to 100\%) as a function of its corresponding epoch in the cosmological rest-frame, plotting in logarithmic-logarithmic space. As an initial result, we discovered that $\pi_{\rm obs}$ tends to be smaller over time. Figure \ref{fig:piTimefull} shows the polarization degree $\pi_{\rm obs}$ as a function of time $t/(1+z)$ after the burst at the cosmological distance for the full sample, where $t$ =($T_{\rm stop}$+$T_{\rm start}$)/2), and $T_{\rm start}$ and $T_{\rm stop}$ are the start time and stop time of the related epoch listed in the Table \ref{tab:fullsample}, and errors $\sigma_{\rm t}$ can be calculated by $\sigma_{\rm t}$=($T_{\rm stop}$-$T_{\rm start}$)/2, and $z$ is reshift. 
Using this full sample, we find that the polarization lightcurve can be best fitted by the power-law model ($\pi=At^{-\alpha}$), the best-fit function gives
\begin{eqnarray}\label{PLfitRest}
\rm log_{10} (\pi_{\rm obs})/(\%)=(4.40\pm 0.36)+(-0.498 \pm 0.020)\times log_{10} [{\it t}/(1+z)]/(s),
\end{eqnarray} 
with the number of data points $N$=115, the Spearman's rank correlation coefficient of $R$=-0.86, and a chance probability $p< 10^{-9}$. These results indicate that the time-sampled polarization evolution over long-timescale periods decays as a power-law\footnote{It's worth noting that the ``$t_0$'' effect might have a significant impact on how early polarization evolves. This happens often in multi-pulse bursts in which adjacent pulses are well-separated by a long quiescent period. The zero time on polarization measurements may need to be corrected to be physical due to the GRB trigger time being no longer special in such cases.} at cosmological distances, which takes the form of $\pi_{\rm obs} \propto [t/(1+z)]^{-0.50 \pm 0.02}$. 

\subsection{Correlation coefficient analysis from the Markov chain Monte Carlo algorithm}

\begin{figure}[ht!]
\includegraphics[width=1\textwidth]{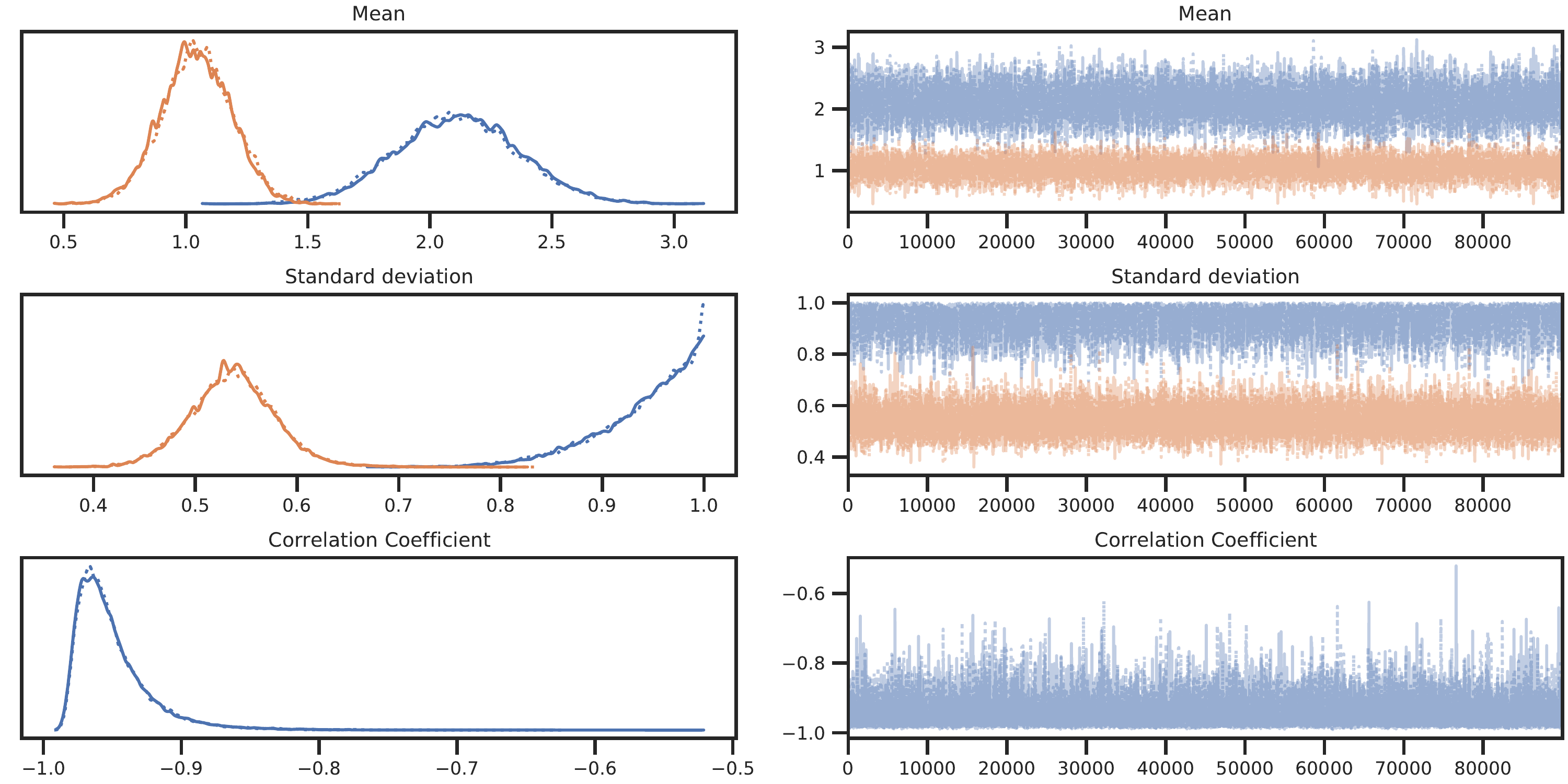}
\caption{MCMC iteration for the mean values and standard deviation of the $t_{\rm z}$ and $\pi$, and their correlation coefficient for our sample, including the value of 10$^{5}$ iterations (right) and their distribution (left).}\label{fig:piTimePymc3}
\end{figure}

Our analysis results were also double-checked using a correlation coefficient analysis from the Markov chain Monte Carlo algorithm. The correlation coefficient can be used to assess the reliability of the correlation between the cosmological rest-frame time $t/(1+z)$ and the GRB observational polarization ($\pi$) for our samples. By implementing the Python package PyMC3, ref.\cite{Salvatier2016}, we use a normal-LKJ correlation prior distribution and the Markov chain Monte Carlo (MCMC) algorithm to obtain the covariance matrix of the multivariate normal distribution, and the correlation coefficient between the parameters is obtained as a result\cite{chib2001markov} by iterating 10$^{5}$ times and burning the first 10$^{4}$ times of the MCMC samples. Our results are summarized in Table \ref{tab:MCMC}, including the expected values ($\mu_1$) and standard deviation ($\sigma_1$) of the normal distribution of the analyzed parameters, and their correlation coefficients, as well as the related Highest Density Interval (HDI) of the posterior distributions, ranging from 3\% to 97\%. In Extended Data Figure \ref{fig:piTimePymc3}, we show an MCMC iteration for the mean values and standard deviation of the $t_{\rm z}$ and $\pi$, and their correlation coefficient, including the value of 10$^{5}$ iterations (right) and their distribution (left).
\begin{table}[ht!]
\refstepcounter{table}\label{tab:MCMC}
{\bf Results of correlation coefficient of Markov chain Monte Carlo algorithm}\\
\begin{tabular}{llllllll}
\hline
Sample&$\mu_1$&$\sigma_1$&$\mu_2$&$\sigma_2$&Correlation Coefficient&hdi interval\\
&($t_{\rm z}$)&($t_{\rm z}$)&($\pi$)&($\pi$)&[3\% to 97\%]\\
\hline
Time-integrated GRB polarization data &2.112$\pm$0.253&0.941$\pm$0.049&1.044$\pm$0.145&0.537$\pm$0.042&-0.950$\pm$0.028&[-0.985 to -0.902]\\
\hline
\end{tabular}
\end{table}

\subsection{``Magnetic patch'' polarization properties and the $\Pi_{\rm max}$ point}\label{sec:Stoke}

After a relativistic jet is emitted, the outflow is likely to carry a fraction of the magnetic field energy in the form of a random magnetic field, known as relativistic magnetized jets. In this random magnetic field, the degree of order of the magnetic field is related to the observed area. The smaller the area is, the higher the degree of order it contains. Naturally, there is a critical area, $S_{\rm patch}$, know as ``magnetic patch\cite{Gruzinov1999,Granot2003,Nakar2004,Granot2005}'', where the magnetic field can be considered to be completely ordered. As shown a yellow arrowed spot in sketch Figure \ref{fig:jet}, we indicate the ``magnetic patch'', where magnetic field orientations are in order, producing polarized photons\footnote{Here, instead of specifying how the magnetic domain is form, we refer such patch to the region where polarized photons are produced in general.}. The magnetic patch area is $S_{\rm patch}$ and its contribution to the observed polarization is about $\Pi_{\rm max}$, where $\Pi_{\rm max}$ is the possibly maximal polarization.

In astrophysical observations, the photon polarization is measured by using the percentages (ranging from 0\% to 100\% percent) of polarized and non-polarized photons to represent the degree of GRB polarization, namely the measured polarization degree is given by  
\begin{equation}
P_{\rm obs}=\pi_{\rm obs}=\frac{I^{\rm p}}{I^{\rm p}+I^{\rm n}},\label{obsp}\\
\end{equation}
where $I^{\rm n}$ and $I^{\rm p}$ respectively refer to the intensities of the natural and fully polarized lights from the source. There are two components: fully polarized photons (e.g., elliptical, circular, and linear polarized light) and non-polarization photons (natural light). The former is completely polarized with $\pi=1$, and the latter is completely non-polarized with $\pi=0$. We illustrate in Figure 1 the observed $\pi_{\rm obs}$ data and their uncertainties.

The measured $\pi_{\rm obs}$ receives all contributions from the area $S_{\rm obs}$ on the jet plane visible to the observer. As discussed soon, the initial $S_{\rm obs}$ is smaller than a magnetic patch area $S_{\rm patch}$. First we introduce the polarization degree of photons emitted from a ``magnetic patch'', correspondingly to the observation one (Eq.\ref{obsp}), 
\begin{equation}
\pi_{\rm patch}=\frac{I^{\rm p}_{\rm patch}}{I_{\rm patch}},\\
\end{equation}
where $I_{\rm patch}=I^{\rm p}_{\rm patch}+I^{\rm n}_{\rm patch}$, and $I^{\rm p}_{\rm patch}$ and $I^{\rm n}_{\rm patch}$ correspond to the intensities of natural and fully polarized lights from a ``magnetic patch''.
It can be seen from Figure \ref{fig:piTimefull} that the time-integrated polarization lightcurve at $S_{\rm obs} \approx S_{\rm patch}$ is a peak, from which we define as $\Pi_{\rm max}$,
\begin{equation}
\Pi_{\rm max}=\pi_{\rm obs} \mid_{S_{\rm patch}\approx S_{\rm obs}} \approx \pi_{\rm patch}.
\label{eq:Pi_max}
\end{equation}
Based on observations (within the uncertainties), the value of $\Pi_{\rm max}$ could be estimated to be between 70 and 100 percent (see Figure \ref{fig:piTimefull}).
Given a GRB source and its surrounding, we assume that magnetic patches have the similar structure of locally coherent domain, thus the $\Pi_{\rm max}$ of different patches are considered to be approximately the same. The reasons are given below. 

When the observed area $S_{\rm obs}$ becomes larger than many magnetic patches, the observed polarization degree $\pi_{\rm obs}$ contributed by photons coming from different magnetic patches. Moreover, all magnetic patches are randomly distributed in entire observed area $S_{\rm obs}$. The net polarization degree $\pi_{\rm obs}$ is expected to be smaller than $\Pi_{\rm max}$ and monochromatically decrease in time, as the number of observed magnetic patches increases\cite{Gruzinov1999,Granot2003,Nakar2004,Granot2005}. Thus the $\Pi_{\rm max}$ (Eq.\ref{eq:Pi_max}) represents a maximal polarization degree in observation. While the net polarization degree $P_{\rm obs}$ contributed by polarised photons from all visible patches will be defined in due course.

\subsection{Magnetic field orderliness and relativistic geometrical effects in GRB polarization}\label{sec:model} 

We follow the discussions in jet-break phenomenon, the observed area $S_{\rm obs}$ (see grey circular area in Figure \ref{fig:jet}) varies in time depending on the bulk Lorentz factor $\Gamma$ of the outflow. The evolution of such relativistic effects on observed polarization may consist of four possible segments depending on whether the observed area ($S_{\rm obs}$) is greater (or less) than the magnetic patch area ($S_{\rm patch}$) and the area ($S_{\rm jet}$) at the break time.

When the Lorentz factor $\Gamma\gg 1$ (early emission phase, such as when the outflow becomes optically transparent), due to a huge $\Gamma$, the observed cone ($\sim 1/\Gamma$) is very small. Only one or a few magnetic patches are seen, depending on magnetic patch size. If only one magnetic patch is seen, the observed area $S_{\rm obs}$ is smaller than and increasing to the magnetic patch area $S_{\rm patch}$ (Segment I, see $\Gamma\sim 300$ in Figure \ref{fig:jet}), we therefore obtain $\pi_{\rm obs}\approx \Pi_{\rm max}$ for $S_{\rm obs}\lesssim S_{\rm patch}$ from (Eqs.\ref{eq:neti}-\ref{eq:finalP}). Indeed, as shown by three data points in the Segment I, the observed photon polarization $\pi_{\rm obs}$ very slightly increases and reach its maximal value. 
However, detail analysis and data need to consider also the time evolution of random magnetic fields in their sizes and strengths for slightly increasing $\Pi_{\rm max}$ value in the Segment I. This is a subject of future study. Nevertheless, using the data at initial time (around 0-10 s), we can infer the maximal value of polarization degree $\Pi_{\rm max} S_{\rm patch}/S_{\rm obs}$ for $S_{\rm patch}\approx S_{\rm obs}$ in $\Gamma \sim 300$ in Figure \ref{fig:jet}.

When the the observed area $S_{\rm obs}$ is larger than $S_{\rm patch}$ (Segment II, see $\Gamma\sim 50$ in Figure \ref{fig:jet}), it covers more than one magnetic patch. The orientations of magnetic fields and photon polarizations vary from one patch to another, see right illustration in Figure \ref{fig:jet}. Suppose that the magnetic patch $S_{\rm patch}$ emits the polarized radiation of total intensity $I_{\rm patch}$ and polarized intensity ${\bf I}^{\rm p}_{\rm patch}$. Here the bold ${\bf I}^{\rm p}_{\rm patch}$ 
represents the orientated polarization intensity of the value $I^{\rm p}_{\rm patch}=|{\bf I}^{\rm p}_{\rm patch}|$. The intensities 
$I_{\rm patch}$ and $I^{\rm p}_{\rm patch}$ are assumed to be approximately equal for all magnetic patches. 
The observed area $S_{\rm obs}\approx \sum_{S_{\rm patch}}S_{\rm patch}$ is smaller than entire emission region. An Earth based observer see the total intensity and net polarization intensity of radiation emitted from all magnetic patches within the observed area $S_{\rm obs}$. Their values can be approximately expressed as
\begin{eqnarray}
I^{\rm p}&= & \sum_{S_{\rm patch}\in S_{\rm obs}} {\bf I}^{\rm p}_{\rm patch} S_{\rm patch}
\approx I^{\rm p}_{\rm patch} S_{\rm patch}^{\rm eff},\label{eq:neti}\\
I&\approx &\sum_{S_{\rm patch}\in S_{\rm obs}} I_{\rm patch} S_{\rm patch}\approx I_{\rm patch} S_{\rm obs}.
\label{eq:faxi}
\end{eqnarray}
Here we introduce an effective patch 
area $S_{\rm patch}^{\rm eff}$ to represent the net polarization observed from different patches.
\begin{eqnarray}
S_{\rm patch}^{\rm eff}&=& \sum_{S_{\rm patch}\in S_{\rm obs}}(-)^p S_{\rm patch} < S_{\rm obs},\label{eq:deeff}
\end{eqnarray}
where the notation $(-)^p$ symbolically indicates 
the randomness of polarization orientations from patches, indicating the destructive (rather than constructive) phenomenon of superposition of light polarizations from different patches. Due to the relativistic beaming effect, the observed area $S_{\rm obs}\propto \Gamma^{-2}$.  
The observational polarization degree $\pi_{\rm obs}$ (Eq.\ref{obsp}) is given by 
\begin{equation}
\begin{split}
\pi_{\rm obs}=\frac{I^{\rm p}}{I}
\approx \Pi_{\rm max} (\frac{S_{\rm patch}^{\rm eff}}{S_{\rm obs}})\propto \Gamma^2,
\end{split}
\label{eq:finalP}
\end{equation}
where $\Pi_{\rm max}=I^{\rm p}_{\rm patch}/I_{\rm patch}$ is the patch maximal polarization degree (Eq.\ref{eq:Pi_max}). We will discuss that in long time scale when $S_{\rm obs}\gg S_{\rm patch}$, the factor $\Pi_{\rm max}S_{\rm patch}^{\rm eff}$ slowly varies compared with $S_{\rm obs}$ variation, leading to $\pi_{\rm obs}\propto S^{-1}_{\rm obs}\propto\Gamma^2$. It is one of the basic formulae of our theoretical model to study the observed polarization degree of GRBs. 

\begin{figure}[ht!]
\includegraphics[width=1.\textwidth]{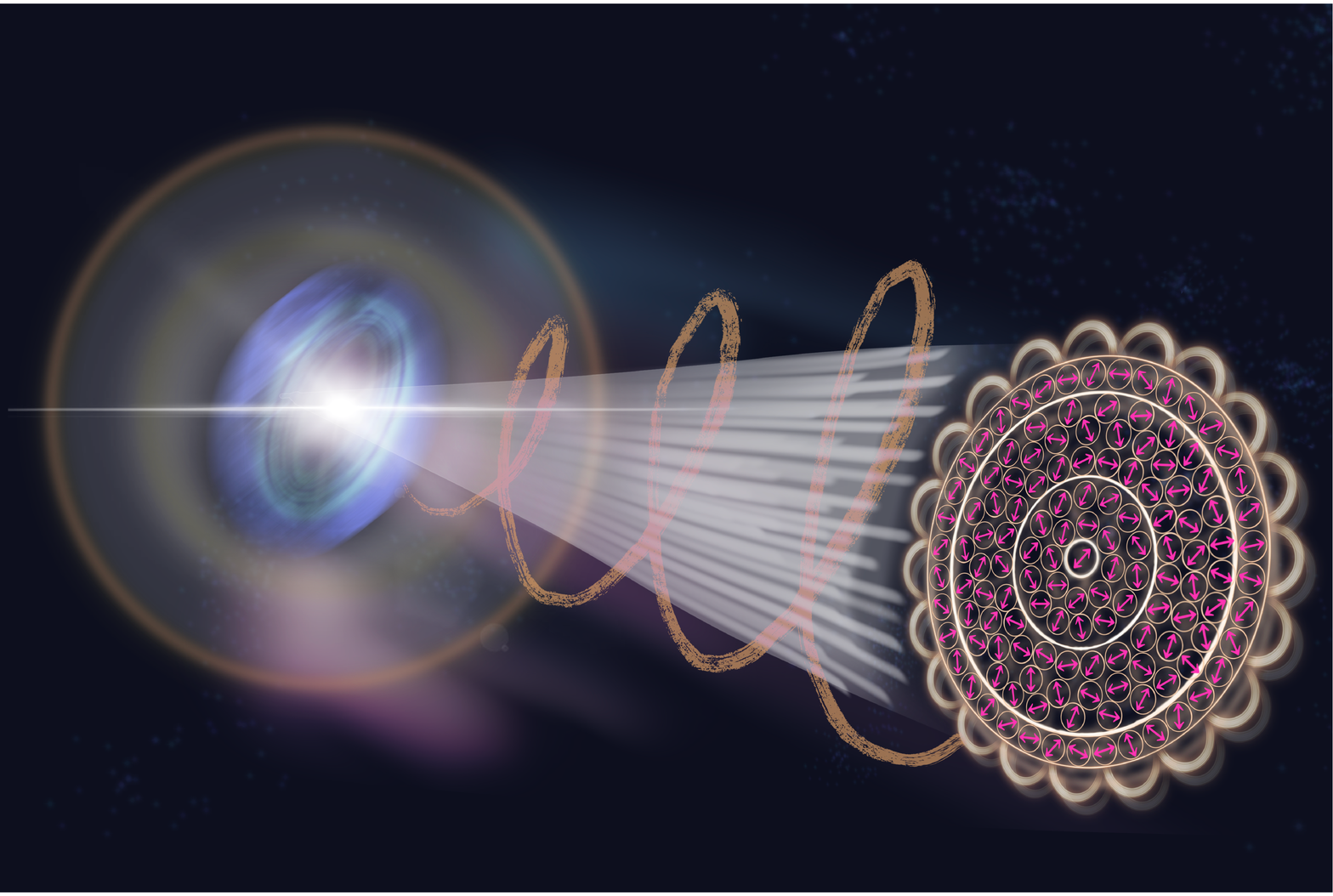}
\caption{A cartoon picture for ``magnetic patches'' on the jet plane represented by many small circles with random red arrows; the observed area $S_{\rm obs}\propto 1/\Gamma^2$ increase in time represented by four hollow circles corresponding to the hollow circles in Fig.\ref{fig:jet}; the irregular jet boundary represented by a waved circle.}
\label{fig:S_patch_catoon}
\end{figure}

Let us consider a GRB jet model\cite{Rhoads1999,Woosley2001} to illustrate the formula (Eq.\ref{eq:finalP}) and its time evolution. According to the jet geometry, at the distance $R_{\rm GRB}\approx ct_{\rm GRB}$ from the center engine and a fixed opening angle $\theta_j$, the jet plane area $S_{\rm jet}\approx \hat{\pi} (\theta_j R_{\rm GRB}/2)^{2}$ 
increasing with time. The magnetic patch area on the jet plane is $S_{\rm patch}\approx \hat{\pi} (\theta_{\rm patch} R_{\rm GRB}/2)^{2}$. We assume that (i) the areas of magnetic patches are approximately equal and (ii) following the dynamics of jet evolution, the magnetic patches are at anchor in the jet flow. Namely, the patch opening angle $\theta_{\rm patch}$ is fixed and the total patch number on the jet plane is approximately constant in time,  
\begin{eqnarray}
N_{\rm patch} &\equiv& \frac{S_{\rm jet}}{S_{\rm patch}}=\frac{\theta^{2}_j}{\theta^{2}_{\rm patch}} \cong {\rm const.}\quad \theta_j> \theta_{\rm patch},
\label{eq:N_patch}
\end{eqnarray}
as well as the patch distribution on the jet plane are assumed to weakly depend on the jet evolution. The observed polarization $\pi_{\rm obs}\approx \Pi_{\rm max}$ is maximal, when $ S_{\rm obs} \approx S_{\rm patch}$. When $ S_{\rm obs} > S_{\rm patch}$, due to the fact that polarization orientations are randomly different among magnetic patches, the net polarization observed is smaller than $\Pi_{\rm max}$. Suppose that the observer is along the line-of-sight of the jet axis, i.e., approximately axial symmetric, and patch magnetic field orientations randomly distribute in the observed area $S_{\rm obs}$ (see Fig. \ref{fig:jet} and Fig. \ref{fig:S_patch_catoon}). The polarizations of radiation from all magnetic patches within the observed area $S_{\rm obs}$ should almost, but not {\it exactly}, cancel each others. The reason is that the number of observed magnetic patches is finite and their orientations cannot be {\it completely} random. To represent the observer relevant net polarization of radiation from different magnetic patches, we introduce an effective patch area $S_{\rm patch}^{\rm eff}$ (Eq.\ref{eq:deeff}). Its value depend on the magnetic patch numbers and their random distribution inside $S_{\rm obs}$. In general, the larger $S_{\rm obs}$ is, the more $S_{\rm patch}$ are seen, the higher degree of their random distribution is, the smaller observed net polarization should be. Therefore, as increasing $S_{\rm obs}$, more and more magnetic patches are observed, the randomness of their magnetic field orientations increases, the effective patch area $S_{\rm patch}^{\rm eff}$ decreases. As a result, the net observed polarization of photons from different magnetic patches decreases. However, as $S_{\rm obs}$ becomes  much larger than $S_{\rm patch}$, the polarization contributions from the patches $S_{\rm patch}$ distributing in the $S_{\rm obs}$ outer area tend to be destructive each other, due to the axial symmetry of their random distribution, as illustrated in sketch Figure \ref{fig:S_patch_catoon}. Namely the observed net polarization contributions mainly come from the patches $S_{\rm patch}$
distributing in the $S_{\rm obs}$ inner area. 
We formally express the $S_{\rm patch}^{\rm eff}$ (Eq.\ref{eq:deeff}) as
\begin{eqnarray}
S_{\rm patch}^{\rm eff}\approx \sum_{\rm inner} (-)^pS_{\rm patch} + \sum_{\rm outer} (-)^pS_{\rm patch} \Rightarrow \sum_{\rm inner} (-)^p S_{\rm patch}.
\label{eq:effratio}
\end{eqnarray}
Its approximate constancy in later time is due to the patch distribution on the jet plane is weakly time dependent, the aforementioned properties of magnetic patches on the jet plane. 
The discussed time evolution of $S_{\rm patch}^{\rm eff}$ is qualitatively sketched in Fig. \ref{fig:S_patch_eff}.

These assumptions are on the basis that the $S_{\rm patch}$ time-varying scale $\tau_{\rm patch}\approx S^{1/2}_{\rm patch}/c$ should be smaller than $S^{1/2}_{\rm obs}/c$. 
In addition, the maximal polarization $\Pi_{\rm max}$ time-varying scale is smaller than $\tau_{\rm patch}$. Therefore, due to the local patch magnetic field dynamics and orientation randomness, we expect in the net polarization the factor $\Pi_{\rm max}S_{\rm patch}^{\rm eff}$ has short-time $\tau_{\rm patch}$ fluctuating features and its mean value weakly depends on time at a much long time scale ($\sim$10$^{5}$ seconds). Thus we parametrize $\Pi_{\rm max}S_{\rm patch}^{\rm eff}\propto t^{-\alpha}$ ($\alpha \ll 1)$, which varies much slowly than $S_{\rm obs}$ in time. As a result, in a long time period of data collection and analysis (Figure \ref{fig:piTimefull}), the time dependence of observed net polarization should be dominantly determined by the time dependence of the observed area $S_{\rm obs}$, whose evolution in time is determined by relativistic effect and kinetic motion of emitting area. 
All above discussions can be generalized to the case that the observer line sign is not exactly along the axial symmetry line of emitting region.

The observed area $S_{\rm obs}$ on the jet plan area $S_{\rm jet}$ evolves in time. In the period of $S_{\rm patch} \lesssim S_{\rm obs} \lesssim S_{\rm jet}$ before the jet-break time scale ($\sim 10^5$ seconds), the observed net polarization mainly follow scaling law $\pi_{\rm obs}\propto 1/S_{\rm obs}$ (Eq.\ref{eq:finalP}) that explains in the numerical data by the following Section (see Methods). We should point out that the local and short-time fluctuating features of the factor $\Pi_{\rm max} S^{\rm eff}_{\rm patch}$ lead to the observed polarization data from GRB sources scattering around the scaling law $\pi_{\rm obs}\propto 1/S_{\rm obs}$, see Figure \ref{fig:jet}. Note that our attention is to explain such a scaling law behavior observed in jet evolution over a long time scale ($\sim 10^5$ seconds). Therefore, we ignore magnetic field and patch fluctuation in small region and short-time scales caused by local and fast varying dynamics, whose averages in a long time and large spatial scales vanish. Moreover, in observed photon polarization samples, these detail features should be also washed away by the randomness of GRB sources in their engines, surroundings and evolution histories. Only GRB common features attributed to the relativistic effect and jet geometry remain, resulting in photon polarization data scattering around the power law in time scale of $10^5$ sec, see Fig. \ref{fig:piTimefull}.

When the jet-break $S_{\rm obs}\approx S_{\rm jet}$ is reached (Segment III, see $\Gamma\sim 10$ in Figure \ref{fig:jet}), we can obtain the value $\pi_{\rm obs}\approx \Pi_{\rm max} S^{\rm eff}_{\rm patch}/S_{\rm jet}$ from data. At this moment, the non-axially symmetric jet boundary is seen, and $S^{\rm eff}_{\rm patch}$ shows a small hump in Fig.\ref{fig:S_patch_eff}.  Such an irregular jet boundary should in principle results in net polarization increases. This explains a slight increasing data in sketch plot in Figure \ref{fig:jet}. 

After the jet-break $S_{\rm obs}> S_{\rm jet}$ (Segment IV, see $\Gamma \gtrsim 1$ in Figure \ref{fig:jet}), the observed net polarization is $\Pi_{\rm max} S^{\rm eff}_{\rm patch}/S_{\rm jet}\propto t^{-\alpha}$, which clearly deviates from the power law. It should remain as an approximate constant soon after jet break, due to the total net polarization of entire jet is observed. Later on, it is expected to slowly decrease in time, namely $\pi_{\rm obs}\propto t^{-\alpha}$, due to the slowly time variations of maximal polarization $\Pi_{\rm max}$, magnetic patch $S_{\rm patch}$ area and distribution, as well as jet dynamics. These features are illustrated by sketches in Figure \ref{fig:jet}. We show in Figure \ref{fig:piTimefull} $\pi_{\rm obs} \propto t^{-0.21 \pm 0.08}$. We note that the index $\alpha\approx 0.21$ is likely to be an upper bound due to only having only one data point after $10^{5}$ seconds of data being used, and need to pending the availability of a sufficiently comprehensive data set. More data analysis at this point are needed, yielding a more definite result. Nevertheless, the index $\alpha\approx 0.30$ being smaller than 0.5 shows that the presented scenario for understanding polarization data is self consistent and contained. In fact, four GRBs (GRB990510, GRB010222, GRB030328, and GRB080928) exhibit the correlation between flux decaying and polarization rising at the same jet break time, which provide evidences for the present theoretical scenario.

\subsection{Correlation between GRB polarization and bulk Lorentz factor}\label{sec:piarea}

Consider a conical jet with line-of-sight aligned with the jet axis and with a jet opening angle $\theta_{j}$, launched from the GRB central engine (see A in Extended Data Figure \ref{fig:S_Gamma}). For simplicity, we assume that the jet opening angle $\theta_{j}$ remains unchanged during the epoch when the jet break happens, corresponding to an area projected onto the plane as $S_{\rm jet}$ (see white circle symbol in the right panel of Figure \ref{fig:jet}), with a radius of $R_{\rm jet}$. Due to relativistic beaming, only the emission inside the 1/$\Gamma$ cone contributes to the observed flux. It projects onto the jet plane an observed area $S_{\rm obs}$ of radius $R_{\rm obs}$ (see orange circle symbol in the right panel of Figure \ref{fig:jet}).

A sketch plot is shown in Extended Data Figure \ref{fig:S_Gamma}, where A is the GRB central engine, B is the observer, O is the center point of the plane, C is the highest latitude photon on the conical jet that the relativistic emission can be observed by the observer, and D is the boundary of the conical jet on the plane. We set
\begin{equation}
\left\{
\begin{array}{ll}
\angle \rm OAC=\alpha,\\
\angle \rm OBC=\beta,\\
\angle \rm BCF=\theta_e,\\
\angle \rm BCE=\varphi=\frac{\theta_e}{2},\\
\angle \rm OAD=\tau=\frac{\theta_j}{2},\\
{\rm OC}=R_{\rm obs},\\
{\rm OD}=R_{\rm jet},\\
{\rm OA}=R_{\rm GRB},\\
{\rm OB}=L_{\rm dis},\\
\end{array} 
\right.
\label{eq:set}
\end{equation}
where $R_{\rm GRB}$ and $L_{\rm dis}$ are the GRB emission radius and luminosity distance, respectively.

Due to relativistic beaming, only the emission inside the 1/$\Gamma$ cone can be attributed to the observed flux. For instance, the relativistic emission produced by the photon C can be observed only inside the cone $\angle$ BCF ($\theta_e$). One therefore has
\begin{equation}
\theta_{e}\simeq\frac{1}{\Gamma}.
\label{eq:???}
\end{equation}
Through geometric correlations, one has
\begin{equation}
\left\{
\begin{array}{ll}
\alpha=\frac{R_{\rm obs}}{R_{\rm GRB}},\\
\beta=\frac{R_{\rm obs}}{L_{\rm dis}}.
\end{array} 
\right.
\end{equation}
On the other hand, due to $L_{\rm dis} \gg R_{\rm GRB}$, 
\begin{equation}
\alpha \gg \beta,
\end{equation}
therefore,
\begin{equation}
\varphi=\alpha+\beta \simeq \alpha,
\end{equation}
and 
\begin{equation}
\theta_e=2\varphi=2\alpha
\end{equation}
\begin{equation}
\left\{
\begin{array}{ll}
R_{\rm jet}=\tau \times R_{\rm GRB}\approx \tau \times ( c t_{\rm GRB}),\\
R_{\rm obs}=\alpha \times R_{\rm GRB} \approx \alpha\times ( c t_{\rm GRB})\\ 
\end{array} 
\right.
\label{eq:RjetRobs}
\end{equation}
where $t_{\rm GRB}$ is the relativistic ejecta emission time measured in the cosmological rest frame, relating to the observed time $t_{\rm obs}$  by $t_{\rm GRB} \simeq 2 \Gamma^{2} t_{\rm obs}$, ref.\cite{Zhang2018}.

As shown in the sketch plot in Figure \ref{fig:jet}, the three characteristic areas are expressed as
\begin{equation}
\left\{
\begin{array}{ll}
S_{\rm jet} = \hat{\pi} (R_{\rm jet})^{2}
=\hat{\pi} (\tau \times R_{\rm GRB})^{2}
=\hat{\pi} (\frac{\theta_j}{2} R_{\rm GRB})^{2}
=\hat{\pi} \frac{\theta^{2}_j}{4}c^{2} t^{2}_{\rm GRB}
=\hat{\pi}(\theta_j c \Gamma^{2} t_{\rm obs})^{2},\\
S_{\rm obs} = \hat{\pi} (R_{\rm obs})^{2}
= \hat{\pi} (\alpha \times R_{\rm GRB})^{2}
= \hat{\pi} (\frac{\theta_e}{2} R_{\rm GRB})^{2}
= \hat{\pi} (\frac{R_{\rm GRB}}{2 \Gamma})^{2}
= \hat{\pi} \frac{c^{2}}{4\Gamma^{2}}t^{2}_{\rm GRB}
=\hat{\pi}(c \Gamma t_{\rm obs})^{2},\\ 
S_{\rm patch} = \hat{\pi} (\frac{\theta_{\rm patch}}{2} R_{\rm GRB})^{2}
= \hat{\pi} \frac{\theta^{2}_{\rm patch}}{4}c^{2} t^{2}_{\rm GRB}
=\hat{\pi}(\theta_{\rm patch} c \Gamma^{2} t_{\rm obs})^{2}.
\end{array} 
\right.
\label{eq:RjetRobs}
\end{equation}

In addition to the constant ratio $N_{\rm patch} \equiv S_{\rm jet}/S_{\rm patch}$ 
(Eq.\ref{eq:N_patch}), we examine below the ratios that are relevant for our description of timing properties of GRB polarization,
\begin{eqnarray}
N_{\rm p} &\equiv& \frac{S_{\rm obs}}{S_{\rm patch}}\approx 
(\theta_{\rm patch}\Gamma)^{-2},
\label{eq:N_p}\\
N_{\rm obs} &\equiv& \frac{S_{\rm jet}}{S_{\rm obs}}=\frac{\theta^{2}_j}{\theta^{2}_e}\approx(\theta_j\Gamma)^{2},
\label{eq:N_obs}
\end{eqnarray}
where $N_{\rm p}$ is the number of patches on the observed area $S_{\rm obs}$ and $N_{\rm obs}$ represents a fraction of $S_{\rm obs}$ over $S_{\rm jet}$. Both $N_{\rm p}$ and 
$N_{\rm obs}$ are $\Gamma$ and time dependent. In terms of the ratio $N_{\rm p}$, the observed polarization degree $\pi_{\rm obs}$ (Eq.\ref{eq:finalP}) can be rewritten as
\begin{eqnarray}
\pi_{\rm obs}= \Pi_{\rm max} (\frac{S_{\rm patch}^{\rm eff}/S_{\rm patch}}{S_{\rm obs}/S_{\rm patch}})=\Pi_{\rm max} (\frac{S_{\rm patch}^{\rm eff}/S_{\rm patch}}{N_{\rm p}}). 
\label{eq:1/N_p}
\end{eqnarray} 
In the initial phase of the Segment II, when $S_{\rm obs} > S_{\rm patch}$ and $S_{\rm obs}$ covers the centre part of the jet plane, the contribution 
$S_{\rm patch}^{\rm eff}/S_{\rm patch}\propto \sqrt{N_{\rm p}}$ following to the probability of random distribution. Thus, the polarization degree $\pi_{\rm obs}\propto 1/\sqrt{N_{\rm p}}$, in consistent with the study of Ref.\cite{Gruzinov1999}. Then from Eq. (\ref{eq:N_p}), we obtain its $\Gamma$-dependence,
\begin{eqnarray}
\pi_{\rm obs}\propto \frac{1}{\sqrt{N_{\rm p}}}\propto \Gamma, 
\label{eq:1/N_p1/2}
\end{eqnarray} 
for large polarization degree in the initial and short period. In the later phase of the Segment II, when $S_{\rm obs}\gg S_{\rm patch}$ and $N_{\rm p}$ increases, the total contribution comes from the inner and outer parts, see (Eq.\ref{eq:effratio}). The outer part contributions cancel among themselves due to an axial symmetry. The inner part contribution remains approximately equal due to the structure and dynamics of the jet and magnetic patch (Eq.\ref{eq:N_patch}). As a result, 
the $N_{\rm p}$- and $\Gamma$-dependence of the polarization degree $\pi_{\rm obs}$ becomes
\begin{eqnarray}
\pi_{\rm obs}\propto \frac{1}{N_{\rm p}}\propto \Gamma^2,
\label{eq:1/N_p1}
\end{eqnarray} 
for small polarization degree in the later and long period. The $\Gamma$-dependence of (Eq.\ref{eq:1/N_p1/2}) and (Eq.\ref{eq:1/N_p1}) is distinctly different. 
Using observed data we examine them in Extended Data Figure \ref{fig:piGamma}.   
It indeed shows that the large polarization degree  follows (Eq.\ref{eq:1/N_p1/2}) (see yellow line in Figure \ref{fig:piGamma}) and the small polarization follows (Eq.\ref{eq:1/N_p1}) (see cyan line in Figure \ref{fig:piGamma}). This supports our theoretical scenario of the time evolution of polarization degree $\pi_{\rm obs}$, described in the previous Section. This is also in accordance with our explanations that the weakly $\Gamma$-depending in the initial Segment I and in the final Segment IV, as well as $\pi_{\rm obs}$ has an increment in the Segment III at the jet break time. 

\subsection{Numerical results (or Self-consistency checking results)}

Using a typical jet opening angle $\theta_j$=0.1, we show the ratios (numbers) $N_{\rm patch}$, $N_{\rm p}$ and $N_{\rm obs}$ as functions of time and Lorentz factor $\Gamma$ in Figs. \ref{fig:Sjet_Sobs_Time} and \ref{fig:Sjet_Sobs_Gamma} respectively. The observed polarization lightcurve shows a hump at initial times 0–10 s (see Segment I in Figure \ref{fig:piTimefull}). 
The peak time occurs at the point $\pi_{\rm obs}\approx \Pi_{\rm max}$ when $S_{\rm obs} \approx S_{\rm patch}$. On the other hand, we have $S_{\rm obs}\approx S_{\rm jet}$ at the jet break time $t_{\rm obs} \simeq$ 3$\times$ 10$^{4}$ seconds (see Segment III in Figure \ref{fig:piTimefull}). At the peak time and jet break time, we numerically calculate the following quantities for consistency checks.

At the initial time $t_{\rm obs} \simeq 3$ seconds and $\Gamma \simeq 300$ for 
a hump $\pi_{\rm obs} \approx \Pi_{\rm max}$, $S_{\rm obs} \approx S_{\rm patch}$, we obtain
\begin{itemize}
\item $R_{\rm GRB} \simeq c t_{\rm GRB}= c (2 \Gamma^{2} t_{\rm obs}) \simeq (3 \times 10^{10} {\rm cm/sec}) \times 2 \times (300)^{2} \times 3 {\rm sec} =1.62 \times 10^{16}$ cm.
\item $S_{\rm obs}\approx S_{\rm patch} =\hat{\pi}(c \Gamma t_{\rm obs})^{2} \simeq 3.14 \times (3 \times 10^{10} {\rm cm/sec})^{2} \times (300)^{2} \times (3 {\rm sec})^{2}=2.5 \times 10^{26} {\rm cm^{2}}$.
\item $S_{\rm jet} =\hat{\pi}(\theta_j c \Gamma^{2} t_{\rm obs})^{2} \simeq 3.14 \times 0.1^{2} \times (3 \times 10^{10} {\rm cm/sec})^{2} \times (300)^{4} \times (3 {\rm sec})^{2}=2.25 \times 10^{28} {\rm cm^{2}}$.
\item $N_{\rm p} \equiv S_{\rm obs}/S_{\rm patch} \approx 1$.
\item $N_{\rm obs} \equiv$ $S_{\rm jet}/S_{\rm obs}\approx\theta^{2}_j\Gamma^{2}=(0.1)^{2} \times (300)^{2}=900$. 
\item $N_{\rm patch}\equiv S_{\rm jet}/S_{\rm patch}\approx S_{\rm jet}/S_{\rm obs}=900$. 
\end{itemize}
At the jet break time $t_{\rm obs} \simeq$ 3$\times$ 10$^{4}$ seconds and $\Gamma \simeq 10$, 
we obtain
\begin{itemize}
\item $R_{\rm GRB} \simeq c t_{\rm GRB}= c (2 \Gamma^{2} t_{\rm obs}) \simeq (3 \times 10^{10} {\rm cm/sec}) \times 2 \times (10)^{2} \times 3 \times 10^{4} {\rm sec}) =1.8 \times 10^{17}$ cm.
\item $S_{\rm obs}\approx S_{\rm jet} =\hat{\pi}(c \Gamma t_{\rm obs})^{2} \simeq 3.14 \times (3 \times 10^{10} {\rm cm/sec})^{2} \times (10)^{2} \times (3 \times 10^{4} {\rm sec})^{2}=2.54 \times 10^{32} {\rm cm^{2}}$.
\item $S_{\rm jet} =\hat{\pi}(\theta_j c \Gamma^{2} t_{\rm obs})^{2} \simeq 3.14 \times (0.1)^{2} \times (3 \times 10^{10} {\rm cm/sec})^{2} \times (10)^{4} \times (3 \times 10^{4} {\rm sec})^{2}=2.54 \times 10^{32} {\rm cm^{2}}$.
\item $N_{\rm p} \equiv S_{\rm obs}/S_{\rm patch} \approx 900$.
\item $N_{\rm obs} \equiv S_{\rm jet}/S_{\rm obs}\approx\theta^{2}_j\Gamma^{2}=(0.1)^{2} \times (10)^{2}=1$. 
\item $N_{\rm patch} \equiv S_{\rm jet}/S_{\rm patch}\approx \frac{\theta^{2}_j}{\theta^{2}_{\rm patch}} =900$. 
\item $\theta_{\rm patch}=0.1/30$.
\end{itemize}
The results are summarized in Table \ref{tab:Seg}.

\subsection{Testing the $\pi_{\rm obs}$-$\Gamma$ and $\pi_{\rm obs}$-[t/(1+z)] relations with observational data (Segment II)}

In the major power-law phase (Segment II, $S_{\rm patch} \lesssim S_{\rm obs} \lesssim S_{\rm jet}$), we have obtained two important scaling laws: the observed net polarization mainly follows a scaling law $\pi_{\rm obs}\propto S^{-1}_{\rm obs}\propto\Gamma^2$ (Eq.\ref{eq:finalP}), and $S_{\rm jet}/S_{\rm obs}=\theta^{2}_j/\theta^{2}_e\propto\Gamma^2$ (Eq.\ref{eq:N_obs}), due to relativistic beaming effect. In Figure \ref{fig:piGamma}, we show that the scaling law $\pi_{\rm obs}\propto\Gamma^2$ is consistent with GRB data.

There are several external-shock models widely discussed in the literature. The simplest model invokes a decelerating blastwave that enters the self-similar regime with constant energy (adiabatic and no energy injection), and constant medium density (ISM)\cite{Meszaros1997, Sari1998}. Several other more complicated cases, such as the dynamics of a blastwave with a varying ambient medium density (e.g., stellar wind\cite{Dai1998,Chevalier1999}), or varying total energy in a blastwave (e.g., with radiative loss\cite{Zhang2018} or with energy injection\cite{Zhang2001}) have also been discussed. These traditional external-shock models give
\begin{equation}
\left\{ \begin{array}{ll}
\Gamma \propto R_{\rm GRB}^{-3/2} \propto t^{-3/8}, R_{\rm GRB} \propto t^{1/4};\quad \rm ISM\\
\Gamma \propto R^{-1/2}_{\rm GRB} \propto t^{-1/4}, R_{\rm GRB} \propto t^{1/2}; \quad \rm wind,\\
\Gamma \propto R^{-3}_{\rm GRB} \propto t^{-3/7}, R_{\rm GRB} \propto t^{1/7}; \quad \rm Radiative\,fireball,\\
\Gamma \propto R^{-\frac{5}{6}}_{\rm GRB} \propto t^{-\frac{5}{16}}, R_{\rm GRB} \propto t^{\frac{3}{8}}; \quad \rm Energy\,injection
\end{array} \right.
\label{eq:ES}
\end{equation}
Note that the energy injection case has taken a typical value of $q$=0.5.

Relativistic effects gives
\begin{equation}
S_{\rm obs} \propto \theta^{2}_e \propto \Gamma^{-2}
\end{equation}

The geometric correlation gives tan($\theta_j$/2)=$R_{\rm jet}/R_{\rm GRB}$, and combining Eq. (\ref{eq:ES}), therefore,
\begin{equation}
S_{\rm jet} \propto R_{\rm GRB}^{2}
\propto \left\{ \begin{array}{ll}
(\Gamma^{-2/3})^{2} \propto \Gamma^{-4/3} \propto t^{1/2},\quad \rm ISM\\
(\Gamma^{-2})^{2} \propto \Gamma^{-4} \propto t^{1}, \quad \rm wind,\\
(\Gamma^{-1/3})^{2} \propto \Gamma^{-2/3} \propto t^{2/7}, \quad \rm Radiative\,fireball,\\
(\Gamma^{-6/5})^{2} \propto \Gamma^{-12/5} \propto t^{3/4}, \quad \rm Energy\,injection
\end{array} \right.
\end{equation}
Before the jet break effect $t \lesssim t_{\rm jet}$, the observed area is $S_{\rm obs}$ on the jet plane and $\pi_{\rm obs} \propto S^{-1}_{\rm obs}$, one therefore has
\begin{equation}
\pi_{\rm obs} \propto S^{-1}_{\rm obs}\propto \Gamma^{2}\propto
\left\{ \begin{array}{ll}
t^{-3/4},\quad \rm ISM\\
t^{-1/2}, \quad \rm wind,\\
t^{-6/7}, \quad \rm Radiative\,fireball,\\
t^{-5/8}, \quad \rm Energy\,injection
\end{array} \right.
\label{eq:piSGamma}
\end{equation}
After the jet break effect $t \gtrsim t_{\rm jet}$, the observed area $S_{\rm obs}$ is larger than the total emission area $S_{\rm jet}$. The total net polarization degree $\pi_{\rm obs}$ has been accounted, thus approximately remains as a constant or slowly decreases in time.

A possible test can be made by investigating the $\pi_{\rm obs}$-$\Gamma(\pi)$ relation via observational data, where $\Gamma(\pi)$ is the bulk Lorentz factor measured during the epoch of polarization. Two main methods have been proposed to estimate $\Gamma$ of a GRB fireball from the literature\cite{Peer2007,Lv2012}, either using a thermal component from prompt emission spectra\cite{Peer2007} or using the early afterglow light curves that show the signal of fireball deceleration\cite{Lv2012}. The former one may not be universally applicable since thermal component spectra have only been observed in a handful of {\it Fermi}-detected GRBs to date. We therefore focus on the latter one. Additionally, $\Gamma(\pi)$ cannot be precisely estimated in many bursts due to their polarization being measured in various emission regions (some in early prompt emission, some others in late after emission). As a proxy of $\Gamma(\pi)$, we use the initial Lorentz factor $\Gamma_{0}$ (at the deceleration radius of the GRBs), which can be estimated by using an empirical relation\cite{Lv2012} ($\Gamma_{0}\simeq 249 L^{0.30}_{\gamma,\rm iso,52}$), where $L_{\gamma,\rm iso}$ is the average isotropic-equivalent luminosity during prompt emission phase. In order to obtain $L_{\gamma,\rm iso}$, we perform time-integrated spectral analysis of the prompt emission for each burst, following the standard practices\cite{Yu2019,Li2021b} used by the full Python package, namely, the Multi-Mission Maximum Likelihood Framework (3ML\cite{Vianello2015}). The energy flux $F_{\gamma}$ (erg cm$^{-2}$s$^{-1}$) can therefore be obtained from the spectral fit, with a $k$-correction ($k_c$) applied. With $F_{\gamma}$ and redshift measurements, one can estimate the average isotropic-equivalent luminosity $L_{\gamma,\rm iso}=4\hat{\pi} d^{2}_{L} F_{\gamma} k_c$.

Extended Data Figure \ref{fig:piGamma} shows the $\pi_{\rm obs}$-$\Gamma_{0}$ correlation. We note that for those bursts in which the polarization is measured during the prompt emission phase, one has $\Gamma_{0}<\Gamma(\pi)$ due to $t_{\Gamma_{0}}<t_{\Gamma(\pi)}$ (see the data points in Extended Data Figure \ref{fig:piGamma} with rightward arrows). While the polarization for the remaining GRBs is measured in the optical band during afterglow emission and has $t_{\Gamma_{0}}>t_{\Gamma(\pi)}$, one therefore has $\Gamma_{0}>\Gamma(\pi)$ (see the data points in Figure \ref{fig:piGamma} with leftward arrows). In summary, the data points are clustered around the solid line defined by the function $\pi_{\rm obs} \propto \Gamma^{2}$ (see the solid line in Figure \ref{fig:piGamma}), which strongly supports our explanations.

We show that the GRB polarization degrees ($\pi_{\rm obs}$) decay as a power-law over time, which takes the form of $\pi_{\rm obs} \propto t^{-0.50 \pm 0.02}$ (Figure \ref{fig:piTimefull}) and the theoretical explanation $\pi_{\rm obs} \propto S^{-1}_{\rm obs}\propto \Gamma^2$ (Eq.\ref{eq:piSGamma}), which is verified by the $\pi_{\rm obs}-\Gamma$ relation (Figure \ref{fig:piGamma}) from data. As a result, using afterglow models of time-decaying Lorentz factors $\Gamma\propto t^{-\alpha/2}$ (Eq.\ref{eq:ES}), we obtain the polarization scaling law $\pi_{\rm obs} \propto t^{-\alpha}$. Therefore, the obtained power-law index of polarization ($\alpha_{\rm obs}$=-0.50 $\pm$ 0.02) from the time-integrated polarization sample can be used as a probe for further information constraints of the afterglow models and environment surrounding GRBs (Extended Data Figure \ref{fig:piTimeModel}).

\subsection{Searching for a jet break in the afterglow lightcurve (Segment III)}

A bit-increment polarization data (deviating from the main power-law function) around $10^{4}$ to $10^{5}$ seconds were observed in 5 GRBs (GRB 990510, GRB 010222, GRB 020405, GRB 030328, and GRB 080928, see Figure \ref{fig:piTimefull}). In the presented scenario, these observed properties are naturally explained as the “jet break” boundary effects that non-axial symmetric irregularities on jet-plan boundary are seen and effective magnetic patch area $S^{\rm eff}_{\rm patch}$ increases in Eq.(\ref{eq:neti}). To verify such an explanation, we further analyse their afterglow lightcurves. This is because the afterglow lightcurves are usually characterized by a post-jet break when the jet boundary is observed, with a steeper temporal decay slope ($\alpha \sim p >1.5$, Ref.\cite{Rhoads1999}) than normal decay, where $p$ is the electron spectral distribution index. As a result shown in the left panel of Extended Data Figure \ref{fig:JetBreakOnsetTime}, it is clearly evident that in these four GRB events, net ploarization increment and afterglow lightcurve decrement are indeed correlated at jet break time.

Extended Data Figure \ref{fig:JetBreakOnsetTime} shows the optical afterglow lightcurves for these five bursts and their polarization data. We fit these lightcurves using a BKPL/SPL model, and pay special attention to their break time (if any) and post-break slope. The best fit results are summarized in Table \ref{tab:JetBreak}. We find that four out of five bursts have post-break slopes larger than 1.5, which can be accounted for by the jet-break phenomenon. In addition, the post-break times measured from the fitting (with uncertainties) of three bursts (GRB 990510, GRB 010222, and GRB 030328) overlap with the epochs of their polarization observations.

{\bf References}
\vspace{1em}


\setcounter{figure}{0}    
\setcounter{section}{0}
\setcounter{table}{0}

\section*{Extended Data}
\begin{figure}[ht!]
\includegraphics[width=1\textwidth]{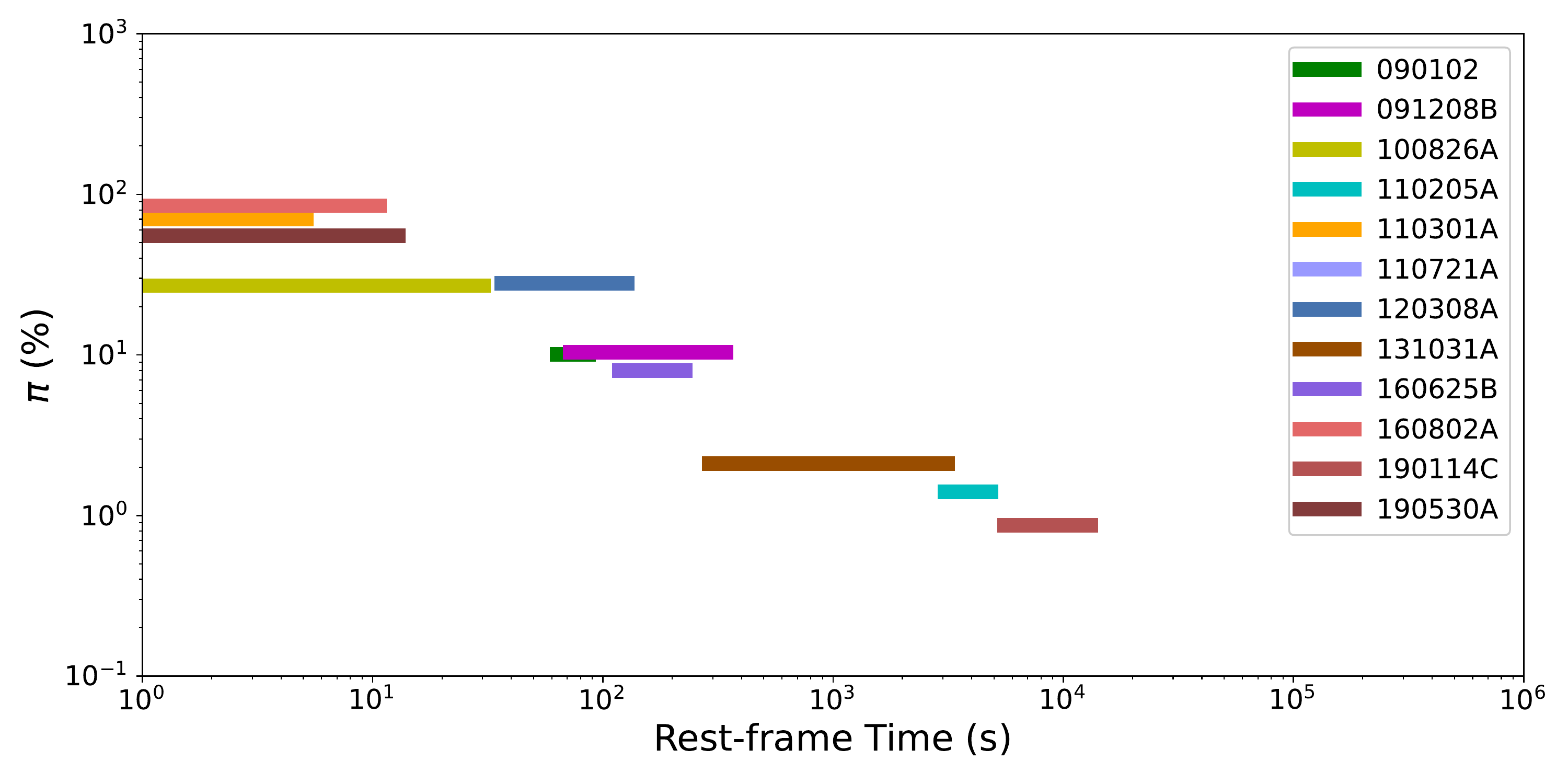}
\caption{The degree of polarization ($\pi_{\rm obs}$) in a time-integrated manner is plotted as a function of their time intervals after the burst at the cosmological distance.}
\label{fig:piTimeInterval}
\end{figure}

\begin{figure*}
\includegraphics[angle=0,scale=0.5]{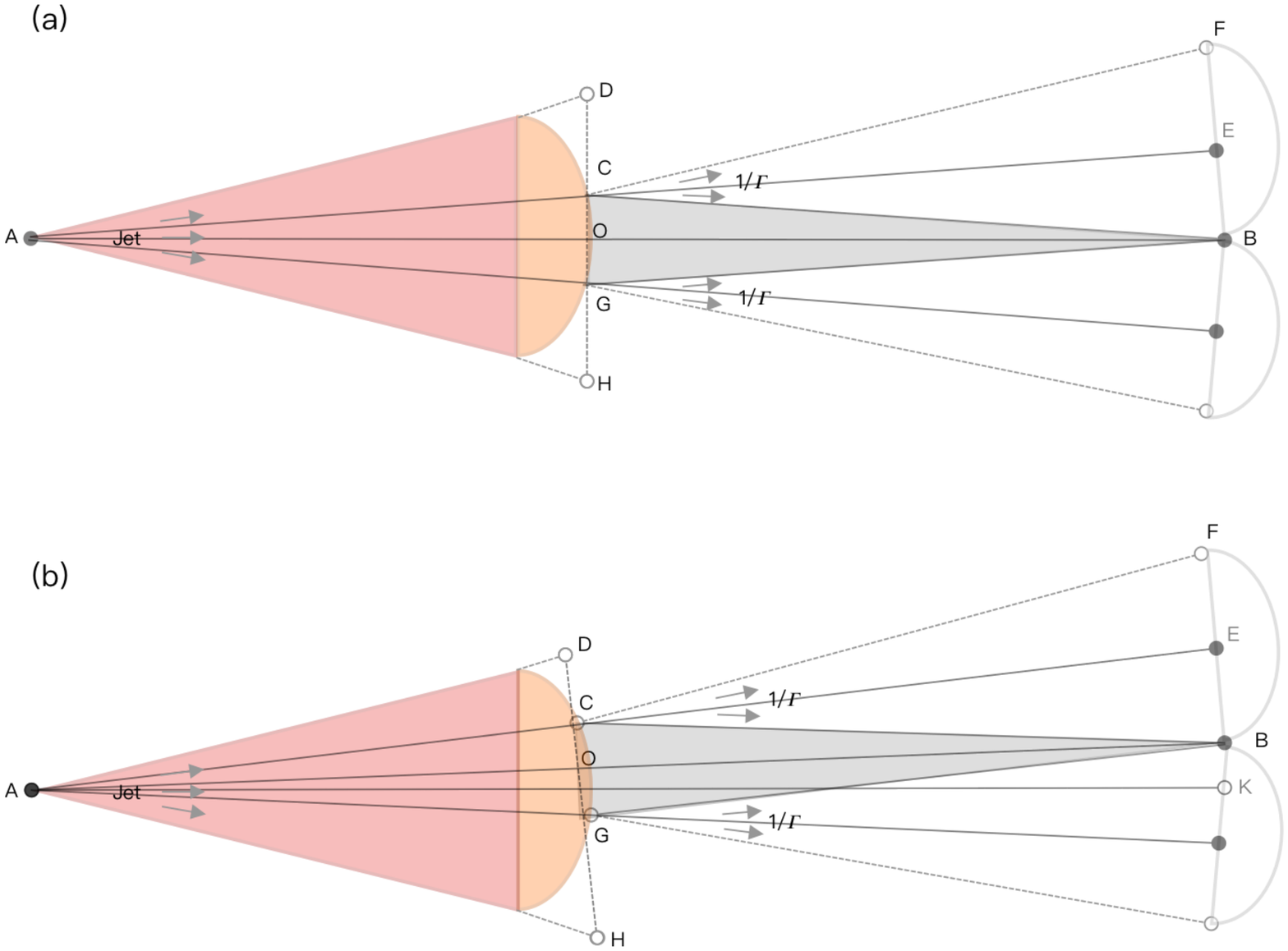}
\caption{Upper panel (a): a sketch plot for delineating the correlation between GRB polarization and the Lorentz factor. We consider a conical jet with line-of-sight aligned with the jet axis with a jet opening angle $\theta_{j}$(=$\angle$HAD). Assuming that A is the GRB central engine, B is the observer, O is the center point of the plane, C(G) are the highest latitude photons on the conical jet that the relativistic emission can be observed by the observer, and D (H) are the boundary of the conical jet on the plane. Due to relativistic beaming, the relativistic emission produced by the photon C can be observed only inside the cone $\theta_e$ ($\angle$ BCF). Lower panel (b): same as (a) but considering a conical jet with line-of-sight unaligned with the jet axis.}\label{fig:S_Gamma}
\end{figure*}

\begin{figure*}
\includegraphics[width=1.0\columnwidth]{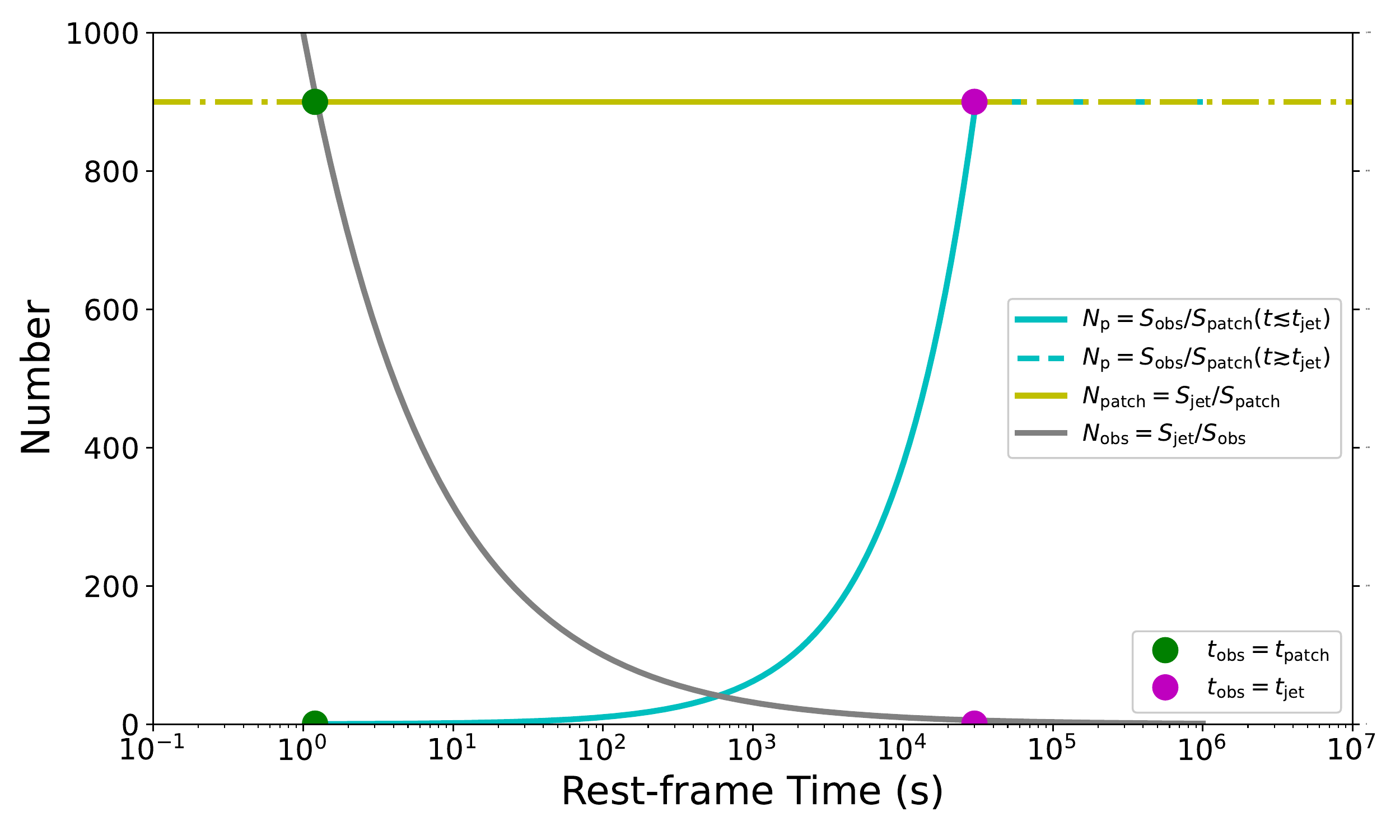}
\caption{The numbers $N_{\rm p}$(=$S_{\rm obs}/S_{\rm patch}$:cyan line), $N_{\rm patch}$(=$S_{\rm jet}/S_{\rm patch}$:yellow line), and $N_{\rm obs}$(=$S_{\rm jet}/S_{\rm obs}$:grey line) as functions of rest-frame time, as well as their typical values (indicated by different colored data points). Data points colored by green and magenta indicate $t_{\rm obs}=t_{\rm patch}$ and $t_{\rm obs}=t_{\rm jet}$, respectively.
$N_{\rm p}$: when $t_{\rm obs} \lesssim t_{\rm patch}$, one has $S_{\rm obs} \lesssim S_{\rm patch}$, and $N_{\rm p} \lesssim 1$; when $t_{\rm patch} \lesssim t_{\rm obs} \lesssim t_{\rm jet}$, one has $S_{\rm patch} \lesssim S_{\rm obs} \lesssim S_{\rm jet}$, and $N_{\rm p} \gtrsim 1$ follows a rapidly rising scaling law over time (cyan line); when $t_{\rm obs} \gtrsim t_{\rm jet}$, one has $S_{\rm obs} \gtrsim S_{\rm jet}$ and $N_{\rm p}$ follows a slowly rising scaling law over time (yellow). $N_{\rm patch}$: compared with $N_{\rm p}$, $N_{\rm patch}$ follows a slowly rising scaling law (note that we have assumed typical observed parameters: $N_{\rm patch}$=900 at the $\Pi_{\rm max}$ point and $N_{\rm patch}$=900 at the jet break time) over time when $t_{\rm obs} \lesssim t_{\rm jet}$. However, $N_{\rm p}$ and $N_{\rm patch}$ will share the same scaling law, when $t_{\rm obs} \gtrsim t_{\rm jet}$. $N_{\rm obs}$: given the typical initial observed parameters (e.g., $N_{\rm obs}$=900, see Methods), $N_{\rm obs}$ as a function of time can be described by the orange line. Note that: one has $N_{\rm obs}\approx 1$, when $t \gtrsim t_{\rm jet}$. Summarized information is given in Table \ref{tab:Seg}.
}\label{fig:Sjet_Sobs_Time}
\end{figure*}

\begin{figure*}
\includegraphics[width=1.0\columnwidth]{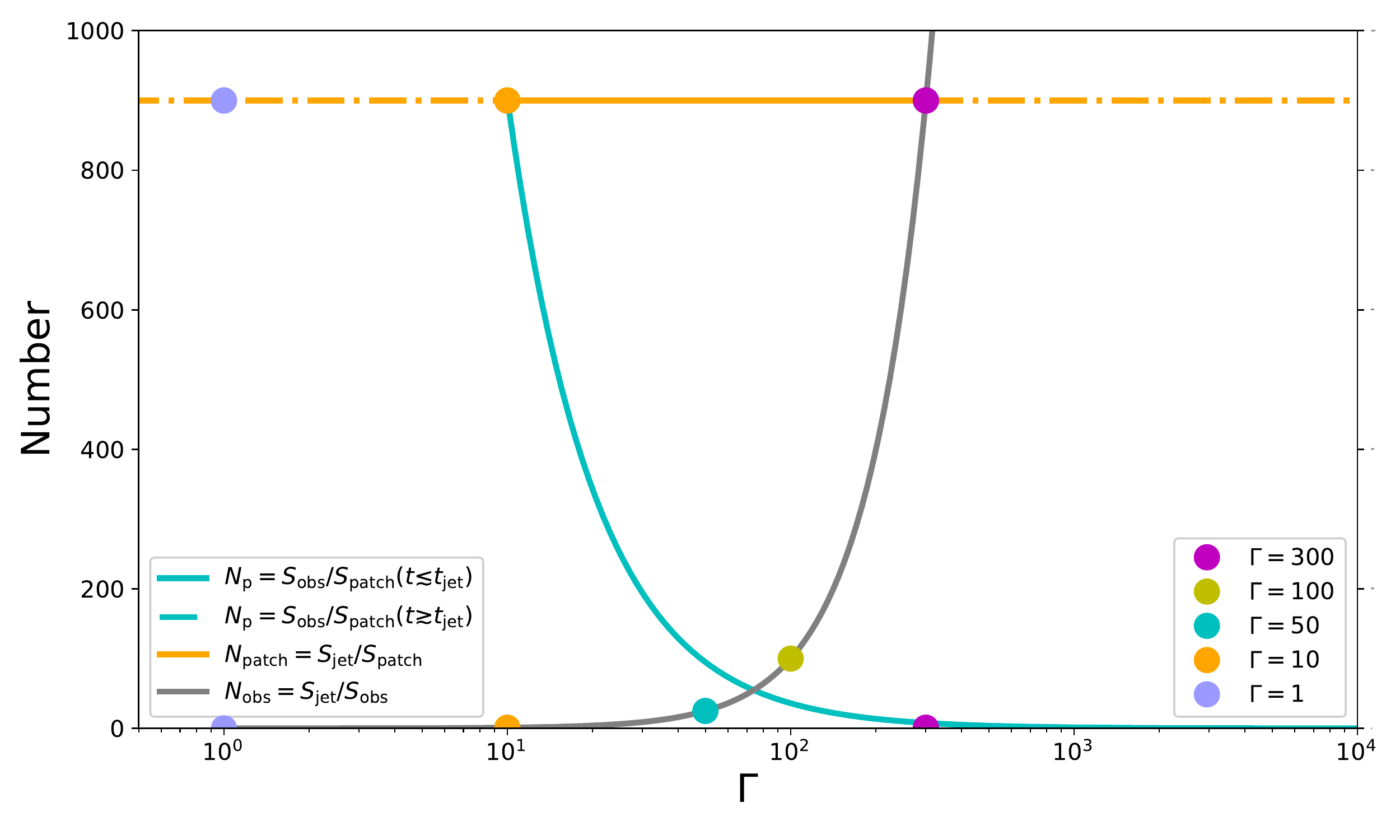}
\caption{The numbers $N_{\rm p}$, $N_{\rm patch}$, and $N_{\rm obs}$ depend on $\Gamma^{2}$. By assuming typical values (e.g., $N_{\rm patch}$=900 when $\Gamma$=300) and a typical jet opening angle ($\theta_j$=0.1) measured from GRBs, $N_{\rm p}$, $N_{\rm patch}$, and $N_{\rm obs}$ as functions of $\Gamma$ can be described by the cyan (Eq.\ref{eq:N_p}), grey (Eq.\ref{eq:N_obs}), and yellow (Eq.\ref{eq:N_patch}) lines. The data points with different colors represent several typical $\Gamma$ values and their corresponding $N$ values. For example, when $\Gamma=300$ (at $\Pi_{\rm max}$ point), one has $N_{\rm p}=1$, $N_{\rm patch}=900$, and $N_{\rm obs}=900$; when $\Gamma=10$ (at jet break time); $N_{\rm p}=900$, $N_{\rm patch}=900$ and $N_{\rm obs}=1$; and when $\Gamma=1$, one has  $N_{\rm p}=900$, $N_{\rm patch}=900$, and $N_{\rm obs}=1$. Note that (i) when $\Gamma=1$, $N_{\rm obs}=1$, rather than $N_{\rm obs}=0.01$ due to the ``jet break'' boundary effect; (ii) when $\Gamma \lesssim 10$ ($t_{\rm obs} \gtrsim t_{\rm jet}$), $N_{\rm p}$ and $N_{\rm patch}$ share the same function (dashed line). Summarized information is given in Table \ref{tab:Seg}.}\label{fig:Sjet_Sobs_Gamma}
\end{figure*}

\begin{figure}[ht!]
\includegraphics[width=1.\textwidth]{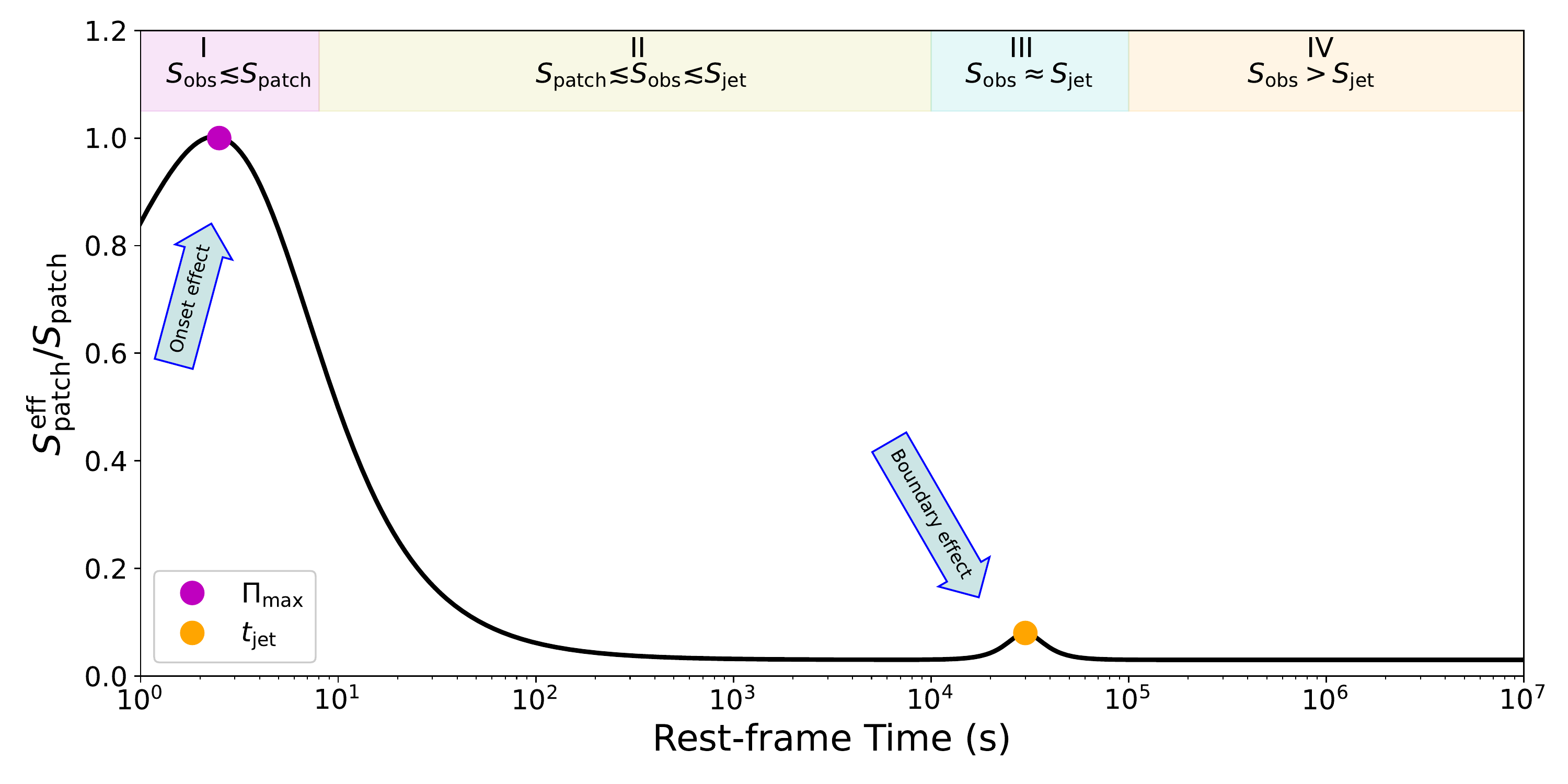}
\caption{The solid line is qualitatively illustrated by the time evolution of $S_{\rm patch}^{\rm eff}/S_{\rm patch}$, while the data points with purple and orange colors indicate the $\Pi_{\rm max}$ point and the typical jet break time.}
\label{fig:S_patch_eff}
\end{figure}

\begin{figure*}
\includegraphics[width=1\columnwidth]{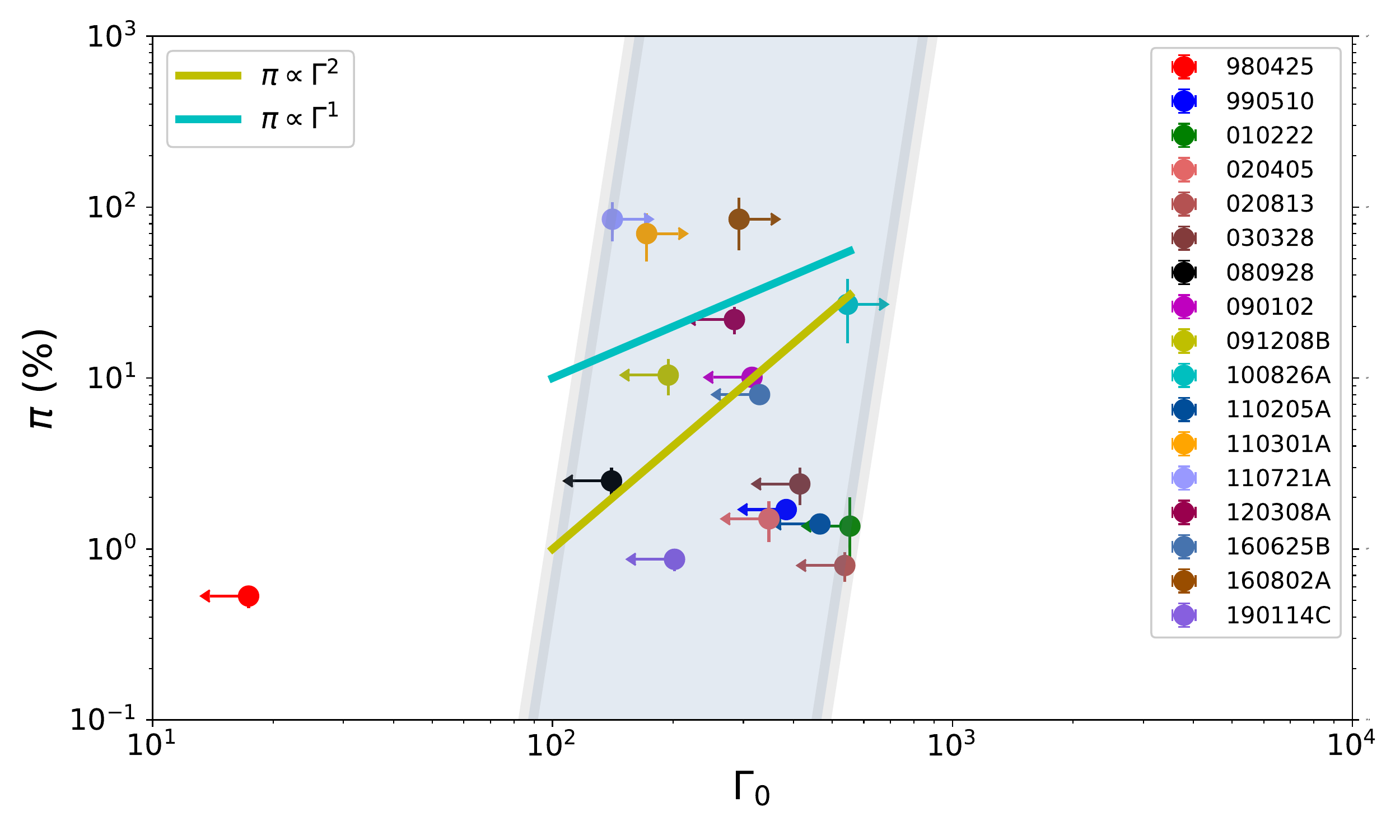}
\caption{Time-integrated degrees of polarization $\pi_{\rm obs}$ as a function of the initial Lorentz factor $\Gamma_{0}$ at the deceleration radius, where $\Gamma_{0}$ is estimated by the empirical relation ($\Gamma_{0}\simeq 249 L^{0.30}_{\gamma,\rm iso,52}$). The cyan solid line represents the function of $\pi_{\rm obs} \propto \Gamma$ (Eq.\ref{eq:1/N_p1/2}) for small polarization degree while the yellow solid line represents the function of $\pi_{\rm obs} \propto \Gamma^{2}$ (Eq.\ref{eq:1/N_p1}) for large polarization degree. The rightward ($\Gamma_{0}<\Gamma(\pi)$: prompt emission phase) and leftward ($\Gamma_{0}>\Gamma(\pi)$: afterglow emission phase) arrows represent the region in which the bulk Lorentz factor is measured during the epoch of polarization.}\label{fig:piGamma}
\end{figure*}

\begin{figure}[ht!]
\includegraphics[width=1\textwidth]{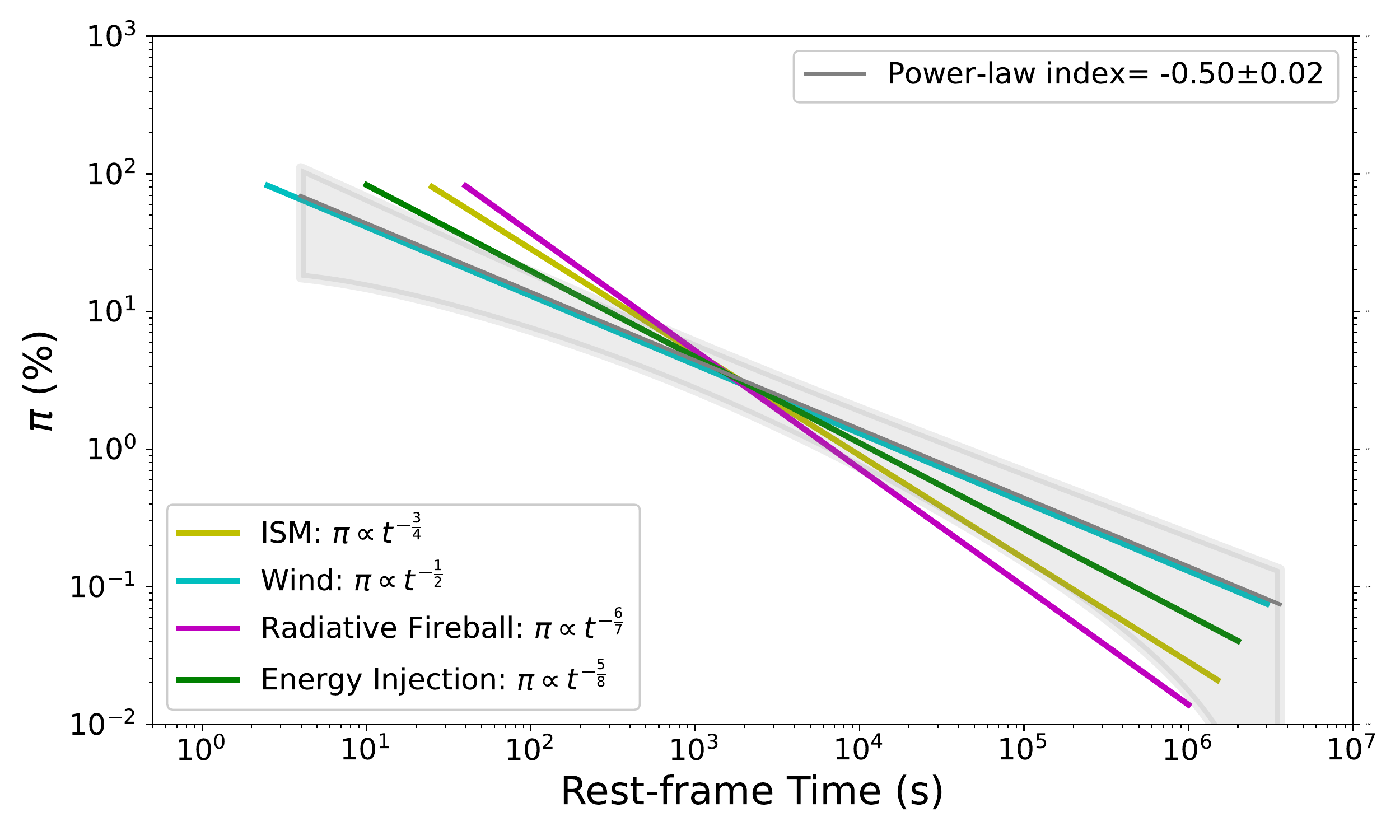}
\caption{Comparing various afterglow models with observed results in the $\pi_{\rm obs}$-[t/(1+z)] plane. The solid line is the best fits using the power-law model from the observational data and with $2\sigma$ (95\% confidence interval) error shadow region. The solid lines indicated by different colors represent different afterglow models.}\label{fig:piTimeModel}
\end{figure}

\begin{figure*}
\includegraphics[width=1.\hsize,clip]{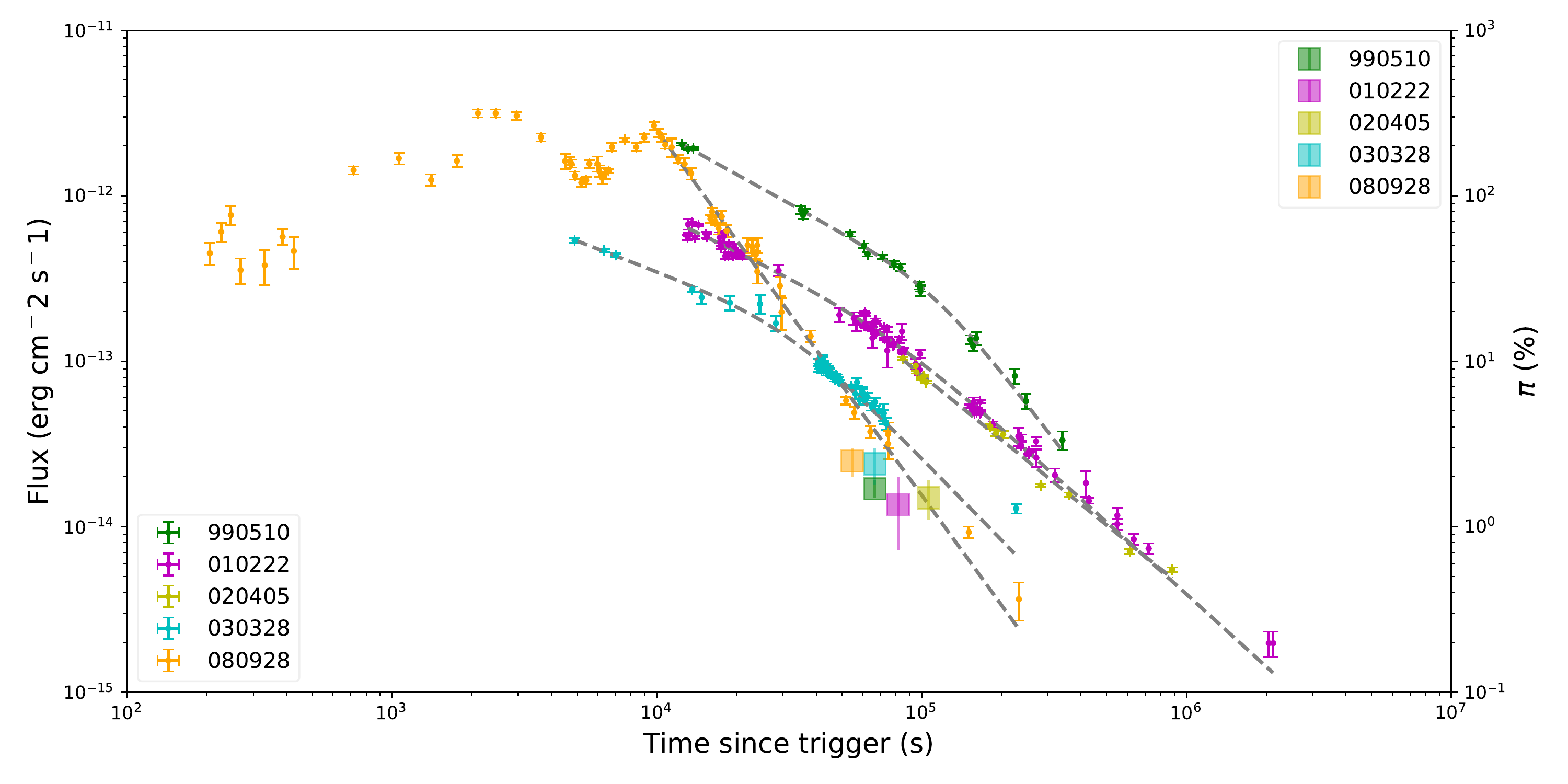}
\caption{Optical ($R$-band) afterglow lightcurves (dotted lines) (left panel) and the degree of polarization (right panel) for the five bursts that have a jet break around the polarization epoch. (GRB 990510, GRB 010222, GRB 020405, GRB 030328, and GRB 080928). The dashed lines in left panel represent the best fits using a BKPL/SPL model.}\label{fig:JetBreakOnsetTime}
\end{figure*}

\begin{enumerate}
\item Extended Data Figure \ref{fig:piTimeInterval}: \textbf{{\bf The degree of polarization ($\pi_{\rm obs}$) in a time-integrated manner is plotted as a function of their time intervals after the burst at the cosmological distance.}}

\item Extended Data Figure \ref{fig:S_Gamma}: \textbf{{\bf Upper panel (a): a sketch plot for delineating the correlation between GRB polarization and the Lorentz factor. We consider a conical jet with line-of-sight aligned with the jet axis with a jet opening angle $\theta_{j}$(=$\angle$HAD). Assuming that A is the GRB central engine, B is the observer, O is the center point of the plane, C(G) are the highest latitude photons on the conical jet that the relativistic emission can be observed by the observer, and D (H) are the boundary of the conical jet on the plane. Due to relativistic beaming, the relativistic emission produced by the photon C can be observed only inside the cone $\theta_e$ ($\angle$ BCF). Lower panel (b): same as (a) but considering a conical jet with line-of-sight unaligned with the jet axis.}}

\item Extended Data Figure \ref{fig:S_patch_catoon}: \textbf{A cartoon picture for ``magnetic patches'' on the jet plane represented by many small circles with random red arrows; the observed area $S_{\rm obs}\propto 1/\Gamma^2$ increase in time represented by four hollow circles corresponding to the hollow circles in Fig.\ref{fig:jet}; the irregular jet boundary represented by a waved circle.}

\item Extended Data Figure \ref{fig:Sjet_Sobs_Time}: \textbf{The numbers $N_{\rm p}$(=$S_{\rm obs}/S_{\rm patch}$:cyan line), $N_{\rm patch}$(=$S_{\rm jet}/S_{\rm patch}$:yellow line), and $N_{\rm obs}$(=$S_{\rm jet}/S_{\rm obs}$:grey line) as functions of rest-frame time, as well as their typical values (indicated by different colored data points). Data points colored by green and magenta indicate $t_{\rm obs}=t_{\rm patch}$ and $t_{\rm obs}=t_{\rm jet}$, respectively.
$N_{\rm p}$: when $t_{\rm obs} \lesssim t_{\rm patch}$, one has $S_{\rm obs} \lesssim S_{\rm patch}$, and $N_{\rm p} \lesssim 1$; when $t_{\rm patch} \lesssim t_{\rm obs} \lesssim t_{\rm jet}$, one has $S_{\rm patch} \lesssim S_{\rm obs} \lesssim S_{\rm jet}$, and $N_{\rm p} \gtrsim 1$ follows a rapidly rising scaling law over time (cyan line); when $t_{\rm obs} \gtrsim t_{\rm jet}$, one has $S_{\rm obs} \gtrsim S_{\rm jet}$ and $N_{\rm p}$ follows a slowly rising scaling law over time (yellow). $N_{\rm patch}$: compared with $N_{\rm p}$, $N_{\rm patch}$ follows a slowly rising scaling law (note that we have assumed typical observed parameters: $N_{\rm patch}$=600 at the $\Pi_{\rm max}$ point and $N_{\rm patch}$=900 at the jet break time) over time when $t_{\rm obs} \lesssim t_{\rm jet}$. However, $N_{\rm p}$ and $N_{\rm patch}$ will share the same scaling law, when $t_{\rm obs} \gtrsim t_{\rm jet}$. $N_{\rm obs}$: given the typical initial observed parameters (e.g., $N_{\rm obs}$=900, see Methods), $N_{\rm obs}$ as a function of time can be described by the orange line. Note that: one has $N_{\rm obs}\approx 1$, when $t \gtrsim t_{\rm jet}$. Summarized information is given in Table \ref{tab:Seg}.}

\item Extended Data Figure \ref{fig:Sjet_Sobs_Gamma}: \textbf{The numbers $N_{\rm p}$, $N_{\rm patch}$, and $N_{\rm obs}$ depend on $\Gamma^{2}$. By assuming typical values (e.g., $N_{\rm patch}$=900 when $\Gamma$=300) and a typical jet opening angle ($\theta_j$=0.1) measured from GRBs, $N_{\rm p}$, $N_{\rm patch}$, and $N_{\rm obs}$ as functions of $\Gamma$ can be described by the cyan (Eq.\ref{eq:N_p}), grey (Eq.\ref{eq:N_obs}), and yellow (Eq.\ref{eq:N_patch}) lines. The data points with different colors represent several typical $\Gamma$ values and their corresponding $N$ values. For example, when $\Gamma=300$ (at $\Pi_{\rm max}$ point), one has $N_{\rm p}=1$, $N_{\rm patch}=900$, and $N_{\rm obs}=900$; when $\Gamma=10$ (at jet break time); $N_{\rm p}=900$, $N_{\rm patch}=900$ and $N_{\rm obs}=1$; and when $\Gamma=1$, one has  $N_{\rm p}=900$, $N_{\rm patch}=900$, and $N_{\rm obs}=1$. Note that (i) when $\Gamma=1$, $N_{\rm obs}=1$, rather than $N_{\rm obs}=0.01$ due to the ``jet break'' boundary effect; (ii) when $\Gamma \lesssim 10$ ($t_{\rm obs} \gtrsim t_{\rm jet}$), $N_{\rm p}$ and $N_{\rm patch}$ share the same function (dashed line). Summarized information is given in Table \ref{tab:Seg}.}

\item Extended Data Figure \ref{fig:S_patch_eff}: \textbf{The solid line is qualitatively illustrated by the time evolution of $S_{\rm patch}^{\rm eff}/S_{\rm patch}$, while the data points with purple and orange colors indicate the $\Pi_{\rm max}$ point and the typical jet break time.}

\item Extended Data Figure \ref{fig:piGamma}: \textbf{Time-integrated degrees of polarization $\pi_{\rm obs}$ as a function of the initial Lorentz factor $\Gamma_{0}$ at the deceleration radius, where $\Gamma_{0}$ is estimated by the empirical relation ($\Gamma_{0}\simeq 249 L^{0.30}_{\gamma,\rm iso,52}$). The solid lines represent the functions of $\pi_{\rm obs} \propto \Gamma^{2}$. The rightward ($\Gamma_{0}<\Gamma_{\pi}$: prompt emission phase) and leftward ($\Gamma_{0}>\Gamma_{\pi}$: afterglow emission phase) arrows represent the region in which the bulk Lorentz factor is measured during the epoch of polarization.}

\item Extended Data Figure \ref{fig:piTimeModel}: \textbf{Comparing various afterglow models with observed results in the $\pi_{\rm obs}$-[t/(1+z)] plane. The solid line is the best fits using the power-law model from the observational data and with $2\sigma$ (95\% confidence interval) error shadow region. The solid lines indicated by different colors represent different afterglow models.}

\item Extended Data Figure \ref{fig:JetBreakOnsetTime}: \textbf{Optical ($R$-band) afterglow lightcurves (dotted lines) (left panel) and the degree of polarization (right panel) for the five bursts that have a jet break around the polarization epoch. (GRB 990510, GRB 010222, GRB 020405, GRB 030328, and GRB 080928). The dashed lines in left panel represent the best fits using a BKPL/SPL model.}

\item Extended Data Table \ref{tab:fullsample}: \textbf{A full catalog of GRB polarimetric observations.}

\item Extended Data Table \ref{tab:JetBreak}: \textbf{Fit Results for the Optical Afterglow Lightcurve (Jet Breaks).} 

\item Extended Data Table \ref{tab:Gamma0}: \textbf{Results of Inferred $\Gamma_{0}$.} 

\end{enumerate}

\clearpage

\clearpage

\clearpage

\clearpage

\clearpage

\clearpage

\clearpage

\clearpage
\setcounter{table}{0}
\begin{table*}
\refstepcounter{table}\label{tab:fullsample}
{\bf Table~1~~A full catalog of GRB polarimetric observations}
\begin{tabular}{llllllllllll}
\hline
GRB&PD&PA&$\nu$/keV&Energy/wave band&Time&Significance&Instrument&Ref.&z\\
&($\pi_{\rm obs}\%$)&($^{\circ}$)&&& ($t-t_{0}$)&&&\\
\hline
930131&$>$35&5&20 keV-1 MeV&$\gamma$-ray&NA&NA&BATSE-CGRO&\cite{Willis2005}&NA\\
\hline
960924&$>$50&47&20 keV-1 MeV&$\gamma$-ray&NA&NA&BATSE-CGRO&\cite{Willis2005}&NA\\
\hline
980329&$<$21&NA&8.3$\times$10$^{9}$&Radio&500h&2$\sigma$&VLA&\cite{Taylor1998}&3.5ph$^{a}$\\
\hline
980425&0.6&80&4000-7000 $\mathring{A}$&Optical&8days&NA&NA&\cite{Patat2001}&0.0085\\
980425&0.4&67&4000-7000 $\mathring{A}$&Optical&25days&NA&NA&\cite{Patat2001}\\
{\bf 980425}&0.53$\pm$0.08&49$\pm$3&390-750 nm&Optical&42days&NA&CTIO 4-m telescope&\cite{Kay1998}\\
\hline
980703&$<$8&NA&4.86$\times$10$^{9}$&Radio&100h&3$\sigma$&VLA&\cite{Frail1998GCN}&0.966\\
980703&$<$8&NA&8.46$\times$10$^{9}$&Radio&100h&3$\sigma$&VLA&\cite{Frail1998GCN}\\ 
\hline
990123&$<$2.3&NA&4.55$\times$10$^{14}$&Optical(R)&18.25h&2$\sigma$&NOT&\cite{Hjorth1999Sci}&1.60\\ 
990123&$<$23$^{l}$,$<$32$^{c}$&NA&8.46$\times$10$^{9}$&Radio&30h&3$\sigma$&VLA&\cite{Kulkarni1999ApJ}\\
\hline
{\bf 990510}&1.7$\pm$0.2&101$\pm$3&4.55$\times$10$^{14}$&Optical(R)&18.5h&NA&VLT&\cite{Covino1999}&1.619\\
990510&$1.6 \pm 0.2$&96$\pm$4&4.55$\times$10$^{14}$&Optical(R)&20.6h&NA&VLT&\cite{Wijers1999}\\
990510&$<$3.9&NA&4.55$\times$10$^{14}$&Optical(R)&43h&NA&VLT&\cite{Wijers1999}\\
990510&2.2$^{+1.1}_{-0.9}$&112$^{+17}_{-15}$&4.55$\times$10$^{14}$&Optical(R)&43.4h&NA&VLT&\cite{Wijers1999}\\ 
\hline
990712&$2.9\pm 0.4$&121.1$\pm$3.5&4.55$\times$10$^{14}$&Optical(R)&10.56h&NA&VLT&\cite{Rol2000}&0.434\\
990712&$1.2\pm 0.4$&116.2$\pm$10.1&4.55$\times$10$^{14}$&Optical(R)&16.8h&NA&VLT&\cite{Rol2000}\\
990712&$2.2\pm 0.7$&139.1$\pm$10.4&4.55$\times$10$^{14}$&Optical(R)&34.8h&NA&VLT&\cite{Rol2000}\\
\hline
991216&$<$2.7&NA&5.45$\times$10$^{14}$&Optical(V)&35.0h&2$\sigma$&VLT&\cite{Covino2004}&1.02?\\ 
991216&$<$5&NA&5.45$\times$10$^{14}$&Optical(V)&60.0h&2$\sigma$&VLT&\cite{Covino2004}\\ 
991216&$<$11$^{l}$,$<$17$^{c}$&NA&8.46$\times$10$^{9}$&Radio&35.8h&3$\sigma$&VLA&\cite{Granot2005}\\
991216&$<$9$^{l}$,$<$15$^{c}$&NA&8.46$\times$10$^{9}$&Radio&64.3h&3$\sigma$&VLA&\cite{Granot2005}\\ 
\hline
000301C&$<$30&NA&4.55$\times$10$^{14}$&Optical(R)&43.2h&NA&VLT&\cite{Stecklum2001}&2.03\\
\hline 
{\bf 010222}&$1.36 \pm 0.64$&NA&5.45$\times$10$^{14}$&Optical(V)&22.65h&NA&NOT&\cite{Bjornsson2002}&1.477\\
\hline
011211&$<$2.7&NA&4.55$\times$10$^{14}$&Optical(R)&35h&3$\sigma$&VLT&\cite{Covino2002}&2.14\\
\hline
020405&$<$11$^{l}$,$<$19$^{c}$&NA&8.46$\times$10$^{9}$&Radio&28.6h&3$\sigma$&VLA&\cite{Granot2005}&0.69\\ 
{\bf 020405}&1.50$\pm$0.40&172$\pm$8&4.55$\times$10$^{14}$&Optical(R)&29.5h&NA&VLT&\cite{Masetti2003}\\ 
020405&9.89$\pm$1.3&179.9$\pm$3.8&5.45$\times$10$^{14}$&Optical(V)&31.7h&NA&MMT&\cite{Bersier2003}\\
020405&1.96$\pm$0.33&154$\pm$5&5.45$\times$10$^{14}$&Optical(V)&52.0h&NA&VLT&\cite{Covino2003}\\ 
020405&1.47$\pm$0.43&168$\pm$9&5.45$\times$10$^{14}$&Optical(V)&76.2h&NA&VLT&\cite{Covino2003}\\ 
\hline
020813&1.8-2.4&153-162&3.26-9.37$\times$10$^{14}$&Optical&4.7-7.9h&NA&Keck&\cite{Barth2003}&1.25\\
{\bf 020813}&0.80$\pm$0.16&144$\pm$6&NA&NA&13.92h&NA&NA&\cite{Covino2002GCN}\\
020813&1.07$\pm$0.22&154.3$\pm$5.9&5.45$\times$10$^{14}$&Optical(V)&21.55h&NA&VLT&\cite{Gorosabel2004}\\
020813&1.42$\pm$0.25&137.0$\pm$4.4&5.45$\times$10$^{14}$&Optical(V)&22.5h&NA&VLT&\cite{Gorosabel2004}\\
020813&1.11$\pm$0.22&150.5$\pm$5.5&5.45$\times$10$^{14}$&Optical(V)&23.41h&NA&VLT&\cite{Gorosabel2004}\\
020813&1.05$\pm$0.23&146.4$\pm$6.2&5.45$\times$10$^{14}$&Optical(V)&24.39h&NA&VLT&\cite{Gorosabel2004}\\
020813&1.43$\pm$0.44&155.8$\pm$8.5&5.45$\times$10$^{14}$&Optical(V)&26.80h&NA&VLT&\cite{Gorosabel2004}\\
020813&1.07$\pm$0.53&163.0$\pm$14.6&5.45$\times$10$^{14}$&Optical(V)&27.34h&NA&VLT&\cite{Gorosabel2004}\\
020813&1.37$\pm$0.49&142.1$\pm$8.9&5.45$\times$10$^{14}$&Optical(V)&27.78h&NA&VLT&\cite{Gorosabel2004}\\
020813&1.26$\pm$0.34&164.7$\pm$7.4&5.45$\times$10$^{14}$&Optical(V)&47.51h&NA&VLT&\cite{Gorosabel2004}\\
020813&0.58$\pm$1.08&13.7$\pm$24.4&5.45$\times$10$^{14}$&Optical(V)&97.29h&NA&VLT&\cite{Gorosabel2004}\\
\hline
021004&1.88$\pm$0.46&189$\pm$7&4.55$\times$10$^{14}$&Optical(R)&8.88h&NA&NOT&\cite{Rol2003}&2.33\\ 
021004&2.24$\pm$0.51&175$\pm$6&4.55$\times$10$^{14}$&Optical(R)&9.12h&NA&NOT&\cite{Rol2003}\\ 
021004&$<$0.60&NA&4.55$\times$10$^{14}$&Optical(R)&9.60h&NA&NOT&\cite{Rol2003}\\ 
021004&0.71$\pm$0.13&140$\pm$5&5.45$\times$10$^{14}$&Optical(V)&16.08h&NA&VLT&\cite{Rol2003}\\
021004&$<$5.0&NA&2.43$\times$10$^{14}$&Optical(J)&10.76h&2$\sigma$&TNG&\cite{Lazzati2003}\\ 
021004&0.51$\pm$0.10&126$\pm$5&5.45$\times$10$^{14}$&Optical(V)&14.62h&NA&VLT&\cite{Lazzati2003}\\ 
021004&0.8-1.7&100-147&3.49-8.57$\times$10$^{14}$&Optical&18.83h&NA&VLT&\cite{Lazzati2003}\\ 
021004&0.43$\pm$0.20&45$\pm$12&5.45$\times$10$^{14}$&Optical(V)&90.7h&NA&VLT&\cite{Lazzati2003}\\  
\hline
021206&$80\pm20$&NA&150-2000 keV&$\gamma$-ray&NA&$\geq$5.7$\sigma$&RHESSI&\cite{Coburn2003}&NA\\
021206&0&NA&NA&NA&NA&NA&NA&\cite{Rutledge2004}\\
021206&$<$4.1&NA&150-2000 keV&$\gamma$-ray&NA&NA&NA&\cite{Rutledge2004}\\
021206&$41^{+57}_{-44}$&NA&150-2000 keV&$\gamma$-ray&NA&NA&RHESSI&\cite{Wigger2004}\\
\hline
030226&$<$1.1&NA&4.55$\times$10$^{14}$&Optical&25.39h&2$\sigma$&VLT&\cite{Klose2004}&1.98\\
\hline
{\bf 030328}&2.4$\pm$0.6&170$\pm$7&5.45$\times$10$^{14}$&Optical&18.5h&NA&VLT&\cite{Maiorano2006}&1.52\\
\hline
\end{tabular}
\end{table*}

\setcounter{table}{1}
\begin{table*}
\setlength{\tabcolsep}{0.35em}
\caption{--- continued}
\begin{tabular}{llllllllll}
\hline
GRB&PD&PA&$\nu$/keV&Energy band&Time&Significance&Instrument&Ref.&z\\
&($\pi_{\rm obs}\%$)&($^{\circ}$)&&& ($t-t_{0}$)&&&\\
\hline
\hline
030329&0.92$\pm$0.10&86.13$\pm$2.43&4.55$\times$10$^{14}$&Optical(R)&12.77h&NA&VLT&\cite{Greiner2003}&0.168\\
030329&0.86$\pm$0.09&86.74$\pm$2.40&4.55$\times$10$^{14}$&Optical(R)&13.18h&NA&VLT&\cite{Greiner2003}\\
030329&0.87$\pm$0.09&88.60$\pm$2.64&4.55$\times$10$^{14}$&Optical(R)&13.61h&NA&VLT&\cite{Greiner2003}\\
030329&0.80$\pm$0.09&91.12$\pm$2.88&3.53-4.55$\times$10$^{14}$&Optical&14.04h&NA&VLT&\cite{Greiner2003}\\
030329&0.66$\pm$0.07&78.52$\pm$2.94&4.55$\times$10$^{14}$&Optical(R)&16.61h&NA&VLT&\cite{Greiner2003}\\
030329&0.66$\pm$0.07&76.69$\pm$2.89&4.55$\times$10$^{14}$&Optical(R)&17.11h&NA&VLT&\cite{Greiner2003}\\
030329&0.56$\pm$0.05&74.37$\pm$3.11&4.55$\times$10$^{14}$&Optical(R)&17.62h&NA&VLT&\cite{Greiner2003}\\
030329&1.10$\pm$0.40&70$\pm$11&4.55$\times$10$^{14}$&Optical(R)&36.72h&NA&CAHA&\cite{Greiner2003}\\
030329&1.37$\pm$0.11&61.65$\pm$2.38&4.55$\times$10$^{14}$&Optical(R)&37.20h&NA&VLT&\cite{Greiner2003}\\
030329&1.50$\pm$0.12&62.29$\pm$2.44&4.55$\times$10$^{14}$&Optical(R)&37.92h&NA&VLT&\cite{Greiner2003}\\
030329&1.07$\pm$0.09&59.41$\pm$2.51&3.53-4.55$\times$10$^{14}$&Optical&40.08h&NA&VLT&\cite{Greiner2003}\\
030329&1.09$\pm$0.08&66.07$\pm$2.45&4.55$\times$10$^{14}$&Optical(R)&40.80h&NA&VLT&\cite{Greiner2003}\\
030329&1.02$\pm$0.08&67.05$\pm$2.60&4.55$\times$10$^{14}$&Optical(R)&41.28h&NA&VLT&\cite{Greiner2003}\\
030329&1.13$\pm$0.08&70.56$\pm$2.51&4.55$\times$10$^{14}$&Optical(R)&41.76h&NA&VLT&\cite{Greiner2003}\\
030329&0.52$\pm$0.06&30.76$\pm$5.04&3.53-4.55$\times$10$^{14}$&Optical&64.32h&NA&VLT&\cite{Greiner2003}\\
030329&0.52$\pm$0.12&12.55$\pm$4.63&4.55$\times$10$^{14}$&Optical(R)&64.80h&NA&VLT&\cite{Greiner2003}\\
030329&0.31$\pm$0.07&24.50$\pm$6.94&4.55$\times$10$^{14}$&Optical(R)&65.28h&NA&VLT&\cite{Greiner2003}\\
030329&0.57$\pm$0.09&52.85$\pm$4.08&4.55$\times$10$^{14}$&Optical(R)&84.96h&NA&VLT&\cite{Greiner2003}\\
030329&0.53$\pm$0.08&57.08$\pm$4.06&4.55$\times$10$^{14}$&Optical(R)&85.44h&NA&VLT&\cite{Greiner2003}\\
030329&0.42$\pm$0.10&62.21$\pm$6.10&4.55$\times$10$^{14}$&Optical(R)&85.92h&NA&VLT&\cite{Greiner2003}\\
030329&1.68$\pm$0.18&66.32$\pm$3.38&4.55$\times$10$^{14}$&Optical(R)&135.84h&NA&NOT&\cite{Greiner2003}\\
030329&2.22$\pm$0.28&75.16$\pm$3.32&4.55$\times$10$^{14}$&Optical(R)&183.36h&NA&VLT&\cite{Greiner2003}\\
030329&1.33$\pm$0.14&70.91$\pm$3.31&4.55$\times$10$^{14}$&Optical(R)&230.16h&NA&VLT&\cite{Greiner2003}\\
030329&2.04$\pm$0.57&1.16$\pm$7.64&4.55$\times$10$^{14}$&Optical(R)&326.40h&NA&VLT&\cite{Greiner2003}\\
030329&0.58$\pm$0.10&42.70$\pm$9.26&4.55$\times$10$^{14}$&Optical(R)&540.00h&NA&VLT&\cite{Greiner2003}\\
030329&1.49$\pm$0.56&99.71$\pm$6.60&4.55$\times$10$^{14}$&Optical(R)&696.00h&NA&VLT&\cite{Greiner2003}\\
030329&1.48$\pm$0.48&25.42$\pm$9.41&4.55$\times$10$^{14}$&Optical(R)&900.00h&NA&VLT&\cite{Greiner2003}\\
030329&2.10$\pm$1.20&54.1$\pm$10.4&4.55$\times$10$^{14}$&Optical(R)&35.19h&NA&CAHA&\cite{Klose2004AA}\\
030329&1.97$\pm$0.48&83.2&4.55$\times$10$^{14}$&Optical(R)&36.49h&NA&IAG-USP&\cite{Magalhaes2003GCN}\\
030329&$<$1&NA&8.4$\times$10$^{9}$&Radio&185.04h&3$\sigma$&VLBA&\cite{Taylor2004}\\
030329&$<$1.8&NA&8.4$\times$10$^{9}$&Radio&1992h&3$\sigma$&VLBA&\cite{Taylor2005}\\
030329&$<$4.7&NA&8.4$\times$10$^{9}$&Radio&5208h&3$\sigma$&VLBA&\cite{Taylor2005}\\
\hline
041219A&98$\pm$33&NA&100-350 keV&$\gamma$-ray&NA&$\sim$2.3$\sigma$&INTEGRAL-SPI&\cite{Kalemci2007}&0.3\\ 
041219A&63$^{+31}_{-30}$,96$\pm$40&70$^{+14}_{-11}$&100-350 keV&$\gamma$-ray&NA&$\sim$2$\sigma$&INTEGRAL-SPI&\cite{McGlynn2007}\\
041219A&$\leq4$ and 43$\pm$25&38$\pm$16&200-800 keV&$\gamma$-ray&first and second peak&$<$2$\sigma$&INTEGRAL-IBIS&\cite{Gotz2009}\\ 
\hline
060418&$<$8&NA&5.08$\times$10$^{14}$&Optical&0.057h&2$\sigma$&LT&\cite{Mundell2007Sci}&1.489\\
\hline
061122&29$^{+25}_{-26}$,$<60$&NA&100-1000 keV&$\gamma$-ray&NA&1$\sigma$&INTEGRAL-SPI&\cite{McGlynn2009}&NA\\ 
061122&$>$33&160$\pm$20&250-800 keV&$\gamma$-ray&NA&90\% confidence&INTEGRAL-IBIS&\cite{Gotz2013}\\ 
\hline
071010&$<$1.3&NA&4.55$\times$10$^{14}$&Optical&21.51h&3$\sigma$&VLT&\cite{Covino2008}&0.98\\ 
\hline
080310&$<2.5$&NA&5.45$\times$10$^{14}$&Optical(V)&87171s&2$\sigma$&VLT&\cite{Littlejohns2012}&2.42\\
080310&$<2.5$&NA&5.45$\times$10$^{14}$&Optical(V)&169501s&2$\sigma$&VLT&\cite{Littlejohns2012}\\
080310&$<2.6$&NA&5.45$\times$10$^{14}$&Optical(V)&253724s&2$\sigma$&VLT&\cite{Littlejohns2012}\\
\hline
{\bf 080928}&$2.5 \pm 0.5$&27$\pm$3&4.25-6.00$\times$10$^{14}$&Optical&15.2h&NA&VLT&\cite{Covino2016}&1.692\\
\hline
{\bf 090102}&$10.1 \pm 1.3$&NA&5.08$\times$10$^{14}$&Optical&160.8-220.8s&$>$3$\sigma$&LT-RINGO&\cite{2009Natur.462..767S}&1.547\\
\hline
091018&$<$0.15$^{\rm c}$&NA&4.55$\times$10$^{14}$&Optical(R)&3.74h&2$\sigma$&VLT&\cite{Wiersema2012}&0.971\\
091018&$<$0.32&NA&4.55$\times$10$^{14}$&Optical(R)&3.17h&1$\sigma$&VLT&\cite{Wiersema2012}\\ 
091018&0.21$\pm$0.31&177.0$\pm$47.5&4.55$\times$10$^{14}$&Optical(R)&4.33h&NA&VLT&\cite{Wiersema2012}\\ 
091018&0.56$\pm$0.27&37.7$\pm$24.7&4.55$\times$10$^{14}$&Optical(R)&4.73h&NA&VLT&\cite{Wiersema2012}\\ 
091018&0.26$\pm$0.31&9.2$\pm$43.7&4.55$\times$10$^{14}$&Optical(R)&5.11h&NA&VLT&\cite{Wiersema2012}\\ 
091018&$<$0.32&NA&4.55$\times$10$^{14}$&Optical(R)&5.53h&NA&VLT&\cite{Wiersema2012}\\ 
091018&1.07$\pm$0.30&179.2$\pm$16.1&4.55$\times$10$^{14}$&Optical(R)&5.91h&NA&VLT&\cite{Wiersema2012}\\ 
091018&0.78$\pm$0.31&3.9$\pm$21.2&4.55$\times$10$^{14}$&Optical(R)&6.33h&NA&VLT&\cite{Wiersema2012}\\ 
091018&0.84$\pm$0.30&171.1$\pm$19.9&4.55$\times$10$^{14}$&Optical(R)&6.70h&NA&VLT&\cite{Wiersema2012}\\ 
091018&2.0$\pm$0.8&10.9$\pm$20.9&1.39$\times$10$^{14}$&Optical(R)&10.34h&NA&VLT&\cite{Wiersema2012}\\ 
091018&1.44$\pm$0.32&2.2$\pm$12.6&4.55$\times$10$^{14}$&Optical(R)&10.92h&NA&VLT&\cite{Wiersema2012}\\
091018&0.94$\pm$0.32&8.0$\pm$18.6&4.55$\times$10$^{14}$&Optical(R)&11.30h&NA&VLT&\cite{Wiersema2012}\\
091018&1.73$\pm$0.36&69.8$\pm$11.7&4.55$\times$10$^{14}$&Optical(R)&27.35h&NA&VLT&\cite{Wiersema2012}\\
091018&3.25$\pm$0.35&57.6$\pm$6.1&4.55$\times$10$^{14}$&Optical(R)&27.73h&NA&VLT&\cite{Wiersema2012}\\
091018&1.99$\pm$0.35&27.6$\pm$10.0&4.55$\times$10$^{14}$&Optical(R)&28.16h&NA&VLT&\cite{Wiersema2012}\\
091018&1.42$\pm$0.36&114.6$\pm$14.0&4.55$\times$10$^{14}$&Optical(R)&28.54h&NA&VLT&\cite{Wiersema2012}\\
091018&0.27$\pm$0.38&101.6$\pm$47.1&4.55$\times$10$^{14}$&Optical(R)&28.97h&NA&VLT&\cite{Wiersema2012}\\
091018&1.00$\pm$0.38&102.2$\pm$20.2&4.55$\times$10$^{14}$&Optical(R)&29.35h&NA&VLT&\cite{Wiersema2012}\\
091018&0.41$\pm$0.34&168.8$\pm$36.6&4.55$\times$10$^{14}$&Optical(R)&32.64h&NA&VLT&\cite{Wiersema2012}\\
091018&0.97$\pm$0.32&32.8$\pm$17.8&4.55$\times$10$^{14}$&Optical(R)&33.40h&NA&VLT&\cite{Wiersema2012}\\
091018&1.08$\pm$0.35&88.7$\pm$17.9&4.55$\times$10$^{14}$&Optical(R)&34.78h&NA&VLT&\cite{Wiersema2012}\\
091018&1.45$\pm$0.37&169.0$\pm$14.3&4.55$\times$10$^{14}$&Optical(R)&55.37h&NA&VLT&\cite{Wiersema2012}\\
\hline
{\bf 091208B}&$10.4 \pm 2.5$&92$\pm$6&4.55$\times$10$^{14}$&Optical(R)&149-706s&NA&Kanata&\cite{2012ApJ...752L...6U}&1.063\\
\hline
\end{tabular}
\end{table*}

\setcounter{table}{1}
\begin{table*}
\setlength{\tabcolsep}{0.35em}
\caption{--- continued}
\begin{tabular}{llllllllllll}
\hline
GRB&PD&PA&$\nu$/keV&Energy band&Time&Significance&Instrument&Ref.&z\\
&($\pi_{\rm obs}\%$)&($^{\circ}$)&&& ($t-t_{0}$)\\
\hline
{\bf 100826A}&$27 \pm 11$&159$\pm$18, 75$\pm$20&20 keV-10 MeV&$\gamma$-ray&0-100s&2.9$\sigma$&IKAROS-GAP&\cite{2011ApJ...743L..30Y}&NA\\
\hline
100906A&$<$10&NA&$\sim$5$\times$10$^{14}$&Optical&$\sim$0.5h&60\%&MASTER&\cite{Gorbovskoy2012}&1.727\\ 
\hline
{\bf 110205A}&1.4&NA&4.55$\times$10$^{14}$&Optical(R)&2.73-4.33hours&NA&CAHA&\cite{2011GCN.11696....1G}&2.22\\
110205A&$<16$&NA&5.08$\times$10$^{14}$&Optical&243s&3$\sigma$&LT-RINGO2&\cite{Cucchiara2011}\\
110205A&$<6.2$&NA&5.08$\times$10$^{14}$&Optical&0.93h&2$\sigma$&LT-RINGO2&\cite{Cucchiara2011}\\
\hline
{\bf 110301A}&$70 \pm 22$&73$\pm$11&10 keV-1 MeV&$\gamma$-ray&0-7s&3.7$\sigma$&IKAROS-GAP&\cite{2012ApJ...758L...1Y}&NA\\
\hline
{\bf 110721A}&$84^{+16}_{-28}$&160$\pm$11&10 keV-1 MeV&$\gamma$-ray&0-11s&3.3$\sigma$&IKAROS-GAP&\cite{2012ApJ...758L...1Y}&0.382\\
\hline
{\bf 120308A}&$28\pm4$ to $16^{+5}_{-4}$&34$\pm$4&5.08$\times$10$^{14}$&Optical&$\sim$200-700s&3$\sigma$&LT-RINGO2&\cite{2013Natur.504..119M}&$<$4.5\\
120308A&$23\pm4$&44$\pm$6&5.08$\times$10$^{14}$&Optical&0.0594h&3$\sigma$&LT-RINGO2&\cite{2013Natur.504..119M}\\
120308A&$17^{+5}_{-4}$&51$\pm$9&5.08$\times$10$^{14}$&Optical&0.0892h&3$\sigma$&LT-RINGO2&\cite{2013Natur.504..119M}\\
120308A&$16^{+7}_{-4}$&40$\pm$10&5.08$\times$10$^{14}$&Optical&0.1189h&3$\sigma$&LT-RINGO2&\cite{2013Natur.504..119M}\\
120308A&$16^{+5}_{-4}$&55$\pm$9&5.08$\times$10$^{14}$&Optical&0.1792h&3$\sigma$&LT-RINGO2&\cite{2013Natur.504..119M}\\
\hline
121011A&$<$15&NA&$\sim$5$\times$10$^{14}$&Optical&$\sim$0.5h&60\% confidence&MASTER&\cite{Pruzhinskaya2014}&NA\\
\hline
121024A&0.61$\pm$0.13$^{\rm c}$&NA&4.55$\times$10$^{14}$&Optical(R)&0.15days&NA&VLT&\cite{2014Natur.509..201W}&2.298\\
121024A&4.09$\pm$0.20&163.7$\pm$2.8&4.55$\times$10$^{14}$&Optical(R)&2.69h&NA&VLT&\cite{2014Natur.509..201W}\\
121024A&4.83$\pm$0.20&160.3$\pm$2.3&4.55$\times$10$^{14}$&Optical(R)&2.96h&NA&VLT&\cite{2014Natur.509..201W}\\
121024A&3.82$\pm$0.20&182.7$\pm$3.0&4.55$\times$10$^{14}$&Optical(R)&4.11h&NA&VLT&\cite{2014Natur.509..201W}\\
121024A&3.12$\pm$0.19&175.3$\pm$3.5&4.55$\times$10$^{14}$&Optical(R)&4.46h&NA&VLT&\cite{2014Natur.509..201W}\\
121024A&3.39$\pm$0.18&178.0$\pm$2.9&4.55$\times$10$^{14}$&Optical(R)&4.84h&NA&VLT&\cite{2014Natur.509..201W}\\
121024A&3.49$\pm$0.18&180.3$\pm$3.0&4.55$\times$10$^{14}$&Optical(R)&5.23h&NA&VLT&\cite{2014Natur.509..201W}\\
121024A&3.20$\pm$0.18&174.5$\pm$3.3&4.55$\times$10$^{14}$&Optical(R)&5.62h&NA&VLT&\cite{2014Natur.509..201W}\\
121024A&0.34$\pm$1.09&51.9$\pm$67.5&4.55$\times$10$^{14}$&Optical(R)&23.45h&NA&VLT&\cite{2014Natur.509..201W}\\
121024A&1.49$\pm$0.78&93.1$\pm$26.6&4.55$\times$10$^{14}$&Optical(R)&26.62h&NA&VLT&\cite{2014Natur.509..201W}\\
121024A&2.66$\pm$0.60&83.0$\pm$12.6&4.55$\times$10$^{14}$&Optical(R)&28.62h&NA&VLT&\cite{2014Natur.509..201W}\\
121024A&0.86$\pm$0.72&101.3$\pm$36.4&4.55$\times$10$^{14}$&Optical(R)&29.39h&NA&VLT&\cite{2014Natur.509..201W}\\
\hline
130427A&$<$3.9$^{l}$,$<$2.7$^{c}$&NA&4.8$\times$10$^{9}$&Radio&36h&3$\sigma$&EVN&\cite{vanderHorst2014}&0.3399\\ 
130427A&$<$7.5$^{l}$,$<$5.7$^{c}$&NA&4.8$\times$10$^{9}$&Radio&60h&3$\sigma$&EVN&\cite{vanderHorst2014}\\
130427A&$<$21$^{l}$,$<$15$^{c}$&NA&4.8$\times$10$^{9}$&Radio&110h&3$\sigma$&EVN&\cite{vanderHorst2014}\\
\hline
{\bf 131030A}&$2.1 \pm 1.6$&27$\pm$22&4.55$\times$10$^{14}$&Optical(R)&665s-2h&NA&Skinakas 1.3m&\cite{2014MNRAS.445L.114K}&1.295\\
\hline
140206A&$>28$, $>48$?&80$\pm$15&15-350 keV&$\gamma$-ray&Second peak($\sim$4-26s)&90\% confidence&INTEGRAL-IBIS&\cite{2014MNRAS.444.2776G}&2.73\\
\hline
140430A&$<22$&NA&5.08$\times$10$^{14}$&Optical&124-424s&3$\sigma$&LT-RINGO3&\cite{2015ApJ...813....1K}&NA\\
\hline
150301B&$>7.6$&NA&5$\times$10$^{14}$&Optical&79-89s&3.2$\sigma$&MASTER&\cite{Gorbovskoy2016}&1.5169\\ 
\hline
151006A&$<84$&NA&100-300 keV&hard X-rays&NA&NA&$AstroSat$-CZTI&\cite{Chattopadhyay2019}&NA\\
\hline
160106A&$68.5\pm24$&$-22.5\pm12$&100-400 keV&hard X-rays&NA&$\gtrsim$3$\sigma$&$AstroSat$-CZTI&\cite{Chattopadhyay2019}&NA\\
\hline
160131A&$94 \pm 31$&$41.2\pm5$&100-300 keV&hard  X-rays&NA&\textbf{$\gtrsim$3$\sigma$}&$AstroSat$-CZTI&\cite{Chattopadhyay2019}&0.972\\
\hline
160325A&$58.75\pm23.5$&$10.9\pm17$&100-300 keV&hard X-rays&NA&$\sim$2.2$\sigma$&$AstroSat$-CZTI&\cite{Chattopadhyay2019}&NA\\
\hline
160509A&$<92$,$96\pm40$&$-28.6\pm11$&100-300 keV&hard X-rays&NA&$\sim$2.5$\sigma$&$AstroSat$-CZTI&\cite{Chattopadhyay2019}&1.17\\
\hline
160530A&$<46$&NA&0.2-5 MeV&soft $\gamma$-ray&NA&90\% confidence&COSI&\cite{Lowell2017}&NA\\
\hline
160607A&$<77$&NA&100-300 keV&hard X-rays&NA&NA&$AstroSat$-CZTI&\cite{Chattopadhyay2019}&NA\\
\hline
160623A&$<46$&NA&100-300 keV&hard X-rays&NA&NA&$AstroSat$-CZTI&\cite{Chattopadhyay2019}&0.367\\
\hline
{\bf 160625B}&8.0$\pm$0.5&NA&NA&Optical&95-360s&\textbf{3$\sigma$}&MASTER-IAC&\cite{2017Natur.547..425T}&1.406\\
\hline
160703A&$<55$&NA&100-300 keV&hard X-rays&NA&NA&$AstroSat$-CZTI&\cite{Chattopadhyay2019}&$<$1.5ph\\
\hline
160802A&85$\pm$29&-36.1$\pm$4.6&200-300 keV&hard X-rays&NA&\textbf{$\geq$3$\sigma$}&$AstroSat$-CZTI&\cite{Chattopadhyay2019}&NA\\
{\bf 160802A}&85$\pm$29&$\sim$-32&200-300 keV&hard X-rays&0-20.34s&\textbf{$\sim$3$\sigma$}&$AstroSat$-CZTI&\cite{2018ApJ...862..154C}\\
\hline
160821A&$48.7\pm14.6$&$-34.0\pm5.0$&100-300 keV&hard X-rays&NA&\textbf{$\geq$3$\sigma$}&$AstroSat$-CZTI&\cite{Chattopadhyay2019}&NA\\
160821A&$21^{+24}_{-19}$&NA&100-300 keV&hard X-rays&115-155s&2.2$\sigma$&$AstroSat$-CZTI&\cite{Sharma2019}\\
160821A&$66^{+26}_{-27}$&NA&100-300 keV&hard X-rays&NA&$\sim$5.3$\sigma$&$AstroSat$-CZTI&\cite{Sharma2019}\\
160821A&$71^{+29}_{-41}$&$110^{+14}_{-15}$&100-300 keV&hard X-rays&115-129s&3.5$\sigma$&$AstroSat$-CZTI&\cite{Sharma2019}\\
160821A&$58^{+29}_{-30}$&$31^{+12}_{-10}$&100-300 keV&hard X-rays&131-139s&4$\sigma$&$AstroSat$-CZTI&\cite{Sharma2019}\\
160821A&$61^{+39}_{-46}$&$110^{+25}_{-26}$&100-300 keV&hard X-rays&142-155s&3.1$\sigma$&$AstroSat$-CZTI&\cite{Sharma2019}\\
160821A&$54\pm21$&-39$\pm$4&100-300 keV&hard X-rays&130-149s&$\geq$3$\sigma$&$AstroSat$-CZTI&\cite{Kole2020}\\
\hline
160910A&$93.7\pm30.92$&$43.5\pm4.0$&100-300 keV&hard X-rays&NA&\textbf{$\geq$3$\sigma$}&$AstroSat$-CZTI&\cite{Chattopadhyay2019}&NA\\
\hline
161203A&$16^{+29}_{-15}$&NA&50-500 keV&hard X-rays&NA&NA&POLAR&\cite{Kole2020}&NA\\
\hline
161217C&$21^{+30}_{-16}$&NA&50-500 keV&hard X-rays&NA&NA&POLAR&\cite{Kole2020}&NA\\
\hline
161218A&$7.0^{+10.7}_{-7.0}$&NA&50-500 keV&hard X-rays&NA&NA&POLAR&\cite{Kole2020}&NA\\
161218A&$<$41\%&NA&50-500 keV&hard X-rays&NA&99\% confidence&POLAR&\cite{Zhang2019}\\
\hline
161218B&$13^{+28}_{-13}$&NA&50-500 keV&hard X-rays&NA&NA&POLAR&\cite{Kole2020}&NA\\
\hline
161229A&$17^{+24}_{-13}$&NA&50-500 keV&hard X-rays&NA&NA&POLAR&\cite{Kole2020}&NA\\
\hline
170101A&$6.3^{+10.8}_{-6.3}$&NA&50-500 keV&hard X-rays&NA&NA&POLAR&\cite{Kole2020}&NA\\
170101A&$<$30\%&NA&50-500 keV&hard X-rays&NA&99\% confidence&POLAR&\cite{Zhang2019}\\
\hline
170101B&$60^{+24}_{-36}$&NA&50-500 keV&hard X-rays&NA&NA&POLAR&\cite{Kole2020}&NA\\
\hline
170114A&$10.1^{+10.5}_{-7.4}$&NA&50-500 keV&hard X-rays&NA&NA&POLAR&\cite{Kole2020}&NA\\
170114A&28$\pm$9$^{c}$&NA&50-500 keV&hard X-rays&NA&NA&POLAR&\cite{Zhang2019}\\
170114A&$<$28\%&NA&50-500 keV&hard X-rays&NA&99\% confidence&POLAR&\cite{Zhang2019}\\
\hline
\end{tabular}
\end{table*}

\setcounter{table}{1}
\begin{table*}
\caption{--- continued}
\begin{tabular}{llllllllllll}
\hline
GRB&PD&PA&$\nu$/keV&Energy band&Time&Significance&Instrument&Ref.&z\\
&($\pi_{\rm obs}\%$)&($^{\circ}$)&&& ($t-t_{0}$)&&&\\
\hline
\hline
170127C&$9.9^{+19.3}_{-8.4}$&NA&50-500 keV&hard X-rays&NA&NA&POLAR&\cite{Kole2020}&NA\\
170127C&$<$68\%&NA&50-500 keV&hard X-rays&NA&99\% confidence&POLAR&\cite{Zhang2019}\\
\hline
170206A&$13.5^{+7.4}_{-8.6}$&NA&50-500 keV&hard X-rays&NA&NA&POLAR&\cite{Kole2020}&NA\\
170206A&$<$31\%&NA&50-500 keV&hard X-rays&NA&99\% confidence&POLAR&\cite{Zhang2019}\\
\hline
170207A&$5.9^{+9.6}_{-5.9}$&NA&50-500 keV&hard X-rays&NA&NA&POLAR&\cite{Kole2020}&NA\\
\hline
170210A&$11.4^{+35.7}_{-9.7}$&NA&50-500 keV&hard X-rays&NA&NA&POLAR&\cite{Kole2020}&NA\\
\hline
170305A&$40^{+25}_{-25}$&NA&50-500 keV&hard X-rays&NA&NA&POLAR&\cite{Kole2020}&NA\\
\hline
170320A&$18^{+32}_{-18}$&NA&50-500 keV&hard X-rays&NA&NA&POLAR&\cite{Kole2020}&NA\\
\hline
170817A&$<$12&NA&2.8 GHz&Radio&244days&99\% confidence&VLA&\cite{Corsi2018}&0.0093\\
\hline
171010A&$\sim$40&variable&100-300keV&hard X-rays&NA&NA&$AstroSat$-CZTI&\cite{Chand2019}&0.3285\\
\hline
171205A&0.27$\pm$0.04&-71.3$\pm$3.3&97.5GHz&Radio&5.187 days&5$\sigma$&ALMA$^{1}$&\cite{Urata2019}&0.0368\\
171205A&0.30$\pm$0.06&-67.9$\pm$4.7&90.5GHz&Radio&5.187 days&5$\sigma$&ALMA&\cite{Urata2019}&\\
171205A&$<$0.32&$<$-78.1&92.4GHz&Radio&5.187 days&5$\sigma$&ALMA&\cite{Urata2019}&\\
171205A&0.35$\pm$0.08&-71.3$\pm$5.5&102.5GHz&Radio&5.187 days&5$\sigma$&ALMA&\cite{Urata2019}&\\
171205A&0.31$\pm$0.06&-58.0$\pm$4.9&104.5GHz&Radio&5.187 days&5$\sigma$&ALMA&\cite{Urata2019}&\\
\hline
{\bf 190114C}&$0.87\pm 0.13$&10$\pm$5&97.5$\times$10$^{9}$&Radio&2.2-5.2h&$\geq$5$\sigma$&ALMA&\cite{Laskar2019}&0.425\\
190114C&$0.60\pm 0.19$&-44$\pm$12&97.5$\times$10$^{9}$&Radio&2.2-5.2h&$\geq$5$\sigma$&ALMA&\cite{Laskar2019}\\
190114C&$7.7\pm1.1$ to 2-4&NA&NA&Optical&52-109s&NA&RINGO3&\cite{Jordana-Mitjans2020}\\
\hline
{\bf 190530A}&$55.43\pm 21.30$&46.74$\pm$4.0&100-300keV&hard X-rays&0-25s&$\leq3\sigma$&AstroSat-CZTI&\cite{Gupta2022}&0.9386\\
190530A&$<64.40$&NA&100-300keV&hard X-rays&7.75-12.25s&95\% confidence&AstroSat-CZTI&\cite{Gupta2022}\\
190530A&$53.95\pm 24.13$&48.17$\pm$6.0&100-300keV&hard X-rays&12.25-25.0s&$\leq3\sigma$&AstroSat-CZTI&\cite{Gupta2022}\\
190530A&$49.99\pm 21.80$&49.61$\pm$6.0&100-300keV&hard X-rays&7.75-25.0s&$\leq3\sigma$&AstroSat-CZTI&\cite{Gupta2022}\\
190530A&$<65.29$&NA&100-300keV&hard X-rays&15.0-19.5s&95\% confidence&AstroSat-CZTI&\cite{Gupta2022}\\
\hline
191016A&$<13.4$&NA&V&Optical&3987-4587s&NA&LT-RINGO3&\cite{Shrestha2021}&NA\\
191016A&5.7$\pm$5.6&82$\pm$26.4&V&Optical&4587-5187s&NA&LT-RINGO3&\cite{Shrestha2021}\\
191016A&$<10.8$&NA&V&Optical&5187-5787s&NA&LT-RINGO3&\cite{Shrestha2021}\\
191016A&$<9.2$&NA&V&Optical&5787-6387s&NA&LT-RINGO3&\cite{Shrestha2021}\\
191016A&$<13.5$&NA&V&Optical&6387-6987s&NA&LT-RINGO3&\cite{Shrestha2021}\\
191016A&$<16.8$&NA&V&Optical&6987-7587s&NA&LT-RINGO3&\cite{Shrestha2021}\\
191016A&$<9.1$&NA&R&Optical&3987-4587s&NA&LT-RINGO3&\cite{Shrestha2021}\\
191016A&11.2$\pm$6.6&90.1$\pm$15.4&R&Optical&4587-5187s&NA&LT-RINGO3&\cite{Shrestha2021}\\
191016A&$<5.5$&NA&R&Optical&5187-5787s&NA&LT-RINGO3&\cite{Shrestha2021}\\
191016A&6.1$\pm$6.0&89.8$\pm$30.5&R&Optical&5787-6387s&NA&LT-RINGO3&\cite{Shrestha2021}\\
191016A&$<12.0$&NA&R&Optical&6387-6987s&NA&LT-RINGO3&\cite{Shrestha2021}\\
191016A&$<11.0$&NA&R&Optical&6987-7587s&NA&LT-RINGO3&\cite{Shrestha2021}\\
191016A&4.7$\pm$4.1&93.16$\pm$22.6&I&Optical&3987-4587s&NA&LT-RINGO3&\cite{Shrestha2021}\\
191016A&$<5.2$&NA&I&Optical&4587-5187s&NA&LT-RINGO3&\cite{Shrestha2021}\\
191016A&$<14.0$&NA&I&Optical&5187-5787s&NA&LT-RINGO3&\cite{Shrestha2021}\\
191016A&14.6$\pm$7.2&100$\pm$12.4&I&Optical&5787-6387s&NA&LT-RINGO3&\cite{Shrestha2021}\\
191016A&$<10.7$&NA&I&Optical&6387-6987s&NA&LT-RINGO3&\cite{Shrestha2021}\\
191016A&$<17.6$&NA&I&Optical&6987-7587s&NA&LT-RINGO3&\cite{Shrestha2021}\\
\hline
\end{tabular}
\begin{flushleft}
\textbf{The ``Note". \\
(a): A photometric redshift is an estimate for the recession velocity of an astronomical object such as a galaxy or quasar, made without measuring its spectrum.\\
(l): Linear polarization measurements.\\
(c): Circular polarization measurements.\\
(1): The Atacama Large Millimeter/submillimeter Array.\\
(d): The Gamma-Ray Burst Polarimeter (GAP) on board the small solar-power-sail demonstrator IKAROS.\\
(e): The purpose-built RINGO2 polarimeter20 on the Liverpool Telescope.\\
(f): Compton Spectrometer and Imager.
} 
\end{flushleft}
\end{table*}

\clearpage
\begin{table}[ht!]
\setlength{\tabcolsep}{0em}
\refstepcounter{table}\label{tab:Seg}
{\bf Extended Data Table~2~~Polarimetric Observations in Different Segments}
\begin{tabular}{llllllllll}
\hline
Segment&observed time $t_{\rm obs}$(s)&observed area ($S_{\rm obs}$)&$\Gamma$&$\pi_{\rm obs}$&$\pi_{\rm obs}$(t)&$\pi_{\rm obs}$($S_{\rm obs},\Gamma$)&$N_{\rm p}$&$N_{\rm patch}$&$N_{\rm obs}$\\
&(in the rest-frame)&(on the jet plane)&&&&&($S_{\rm obs}/S_{\rm patch}$)&($S_{\rm jet}/S_{\rm patch}$)&($S_{\rm jet}/S_{\rm obs}$)\\
\hline
I&$\lesssim t_{\rm patch}$($0\sim 3$)&$\lesssim S_{\rm patch}$&$> 300$&$\lesssim \Pi_{\rm max}$&increase&$\propto 1/S_{\rm obs}, \propto \Gamma^{2}$&$\lesssim$1&$\approx$900&$>$900\\
$\Pi_{\rm max}$&$\approx t_{\rm patch}$($\sim$3)&$\approx S_{\rm patch}$&$\approx 300$&$\approx \Pi_{\rm max}$&$\propto t^{-0.50 \pm 0.02}$&$\propto 1/S_{\rm obs}, \propto \Gamma^{2}$&$\approx$ 1&$\approx$900&$\approx$900\\
II&$t_{\rm patch} \lesssim t_{\rm obs}\lesssim t_{\rm jet}$ ($3 \sim 3 \times 10^{4}$)&$S_{\rm patch} \lesssim S_{\rm obs} \lesssim S_{\rm jet}$&$\approx 50$&$< \Pi_{\rm max}$&$\propto t^{-0.50 \pm 0.02}$&$\propto 1/S_{\rm obs}, \propto \Gamma^{2}$&$>$1&$\approx$900&$\approx$25\\
III&$\approx t_{\rm jet}$($10^{4}\sim 10^{5}$)&$\approx S_{\rm jet}$&$\approx 10$&$<\Pi_{\rm max}$&increase&$\propto 1/S_{\rm obs}, \propto \Gamma^{2}$&$\approx$900&$\approx$900&$\approx$1\\
IV&$>t_{\rm jet}$($>10^{5}$)&$> S_{\rm jet}$&$\approx 1$&$< \Pi_{\rm max}$&$\propto t^{-0.21 \pm 0.08}$&$\propto 1/S_{\rm obs}, \propto \Gamma^{2}$&$\approx$900&$\approx$900&$\approx$1\\
\hline
\end{tabular}
\caption*{}
\end{table}

\clearpage
\begin{table}[ht!]
\refstepcounter{table}\label{tab:JetBreak}{\bf Extended Data Table~3~~Fit Results for the Optical Afterglow Lightcurve (Jet Breaks)}\\
\begin{tabular}{llllllll}
\hline
GRB&$t_{\rm start}$$\sim$$t_{\rm stop}$&Model&$\alpha_1$&$t_{\rm break}$&$\alpha_2$&Is a jet break?\\
&($10^{4}$~s)&&&($10^{4}$~s)\\
\hline
990510(R)&1.24$\sim$34.02&BKPL&0.88$\pm$0.02&$11.4\pm1.3$&2.11$\pm$0.30&Yes\\
010222(R)&1.31$\sim$212.48&BKPL&0.67$\pm$0.15&$12.3\pm9.3$&1.86$\pm$0.48&Likely\\
020405(R)&8.50$\sim$88.26&SPL&...&...&1.30$\pm$0.05&No\\
030328(R)&0.49$\sim$22.75&BKPL&0.54$\pm$0.06&4.81$\pm$0.89&2.20$\pm$0.31&Yes\\
080928($R_{\rm c}$)&0.02$\sim$23.30&SPL&...&...&2.21$\pm$0.06 (Late part)&Yes\\
\hline
\end{tabular}
\end{table}

\clearpage
\begin{table}[ht!]
\refstepcounter{table}\label{tab:Gamma0}
\setlength{\tabcolsep}{0.35em}
{\bf Extended Data Table~4~~Results of Inferred $\Gamma_{0}$}
\begin{tabular}{lllllllll}
\hline
GRB&$t_{1}$$\sim$$t_{2}$&$S$ &Detectors
&$\Delta T_{\rm src}$
&$[\Delta T_{\rm (bkg,1)},\Delta T_{\rm (bkg,2)}]$&$F^{\rm obs}_{\gamma}$&$L_{\gamma}$&$\Gamma_{0}$\\
&(s)&&&(s)&(s)&(erg cm$^{-2}$ s$^{-1}$)&(erg s$^{-1}$)\\
\hline
080928&-1$\sim$14.336&11.67&(n3)n6n7b0&(-1 to 14.336)&(-20 to -10, 40 to 60) &$(6.13\pm3.98)\times10^{-8}$&$(1.48\pm0.96)\times10^{51}$&140\\
090102&-1$\sim$26.624&44.69&n9na(nb)b1&(-1 to 26.624)&(-20 to -10, 180 to 200) &$(1.34\pm0.08)\times10^{-6}$&$(2.19\pm0.13)\times10^{52}$&315\\
091208B&-1$\sim$12.480&32.93&n9(na)b0&(-1 to 12.48)&(-20 to -10, 40 to 60) &$(6.98\pm0.17)\times10^{-7}$&$(4.39\pm1.07)\times10^{51}$&195\\
100826A&-1$\sim$82&11.67&n7(n8)b1&(-1 to 82)&(-20 to -10, 200 to 250) &$(3.92\pm0.25)\times10^{-6}$&$(1.37\pm0.09)\times10^{53}$&546\\
110301A&-1$\sim$5.693&269.92&n7(n8)nbb1&(-1 to 5.693)&(-20 to -10, 40 to 60) &$(6.11\pm0.30)\times10^{-6}$&$(2.96\pm0.14)\times10^{51}$&173\\
110721A&-1$\sim$21.822&84.67&(n6)n7n9b1&(-1 to 21.822)&(-20 to -10, 40 to 60) &$(3.76\pm0.40)\times10^{-6}$&$(1.51\pm0.16)\times10^{51}$&141\\
160625B&-1$\sim$300&156.58&(n6)n7n9b1&(-1 to 300)&(-50 to -20, 80 to 100) &$(3.43\pm0.35)\times10^{-6}$&$(2.54\pm0.26)\times10^{52}$&329\\
160802A&-1$\sim$20.34&151.65&(n2)b0&(-1 to 20.34)&(-20 to -10, 60 to 80) &$(4.87\pm0.49)\times10^{-6}$&$(1.71\pm0.17)\times10^{52}$&293\\
190114C&-1$\sim$116.354&189.04&n3(n4)b0&(-1 to 116.354)&(-20 to -10, 180 to 200) &$(7.52\pm0.33)\times10^{-6}$&$(4.96\pm0.22)\times10^{51}$&202\\
\hline
\end{tabular}
\caption*{}
\end{table}

\end{document}